CRANFIELD UNIVERSITY

SAIF SULAIMAN SAIF AL WAHAIBI

DEVELOPMENT OF A COBALT ELECTROCHEMICAL SENSOR FOR MEASURING PHOSPHATE IN MUNICIPAL WASTEWATERS

SCHOOL OF WATER, ENERGY AND ENVIRONMENT
Water and Wastewater Engineering

MSc
Academic Year: 2016 - 2017

Supervisor: Dr. Ana Soares
September 2017

CRANFIELD UNIVERSITY

SCHOOL OF WATER, ENERGY AND ENVIRONMENT
Water and Wastewater Engineering

MSc

Academic Year 2016 - 2017

SAIF SULAIMAN SAIF AL WAHAIBI

Development of a cobalt electrochemical sensor for measuring phosphate in municipal wastewaters

Supervisor: Dr. Ana Soares
September 2017

This thesis is submitted in partial fulfilment of the requirements for the degree of Master of Science

# ABSTRACT

The introduction of the Water Framework directive sets stringent limits on phosphorous discharge from wastewater treatment plants to maintain the complex interdependent relationship between water tributaries and the ecosystem. A critical step in doing so is the accurate and swift measurement of phosphates in wastewater. This paper studies a cobalt based electrochemical sensor for phosphate detection in wastewater. An evaluation of the sensor's operational envelope, impact of pH, detection limits, linearity of response, accuracy and reproducibility in a single ion solution was conducted. An indirect method was employed to assess the effect of all of these parameters; the parameter was kept constant, while the phosphate concertation was varied. Results revealed a dynamic linear range between 6.00 – 100 mg P/L with a detection limit of (6 ± 8) mg P/L at pH 8.00. pH, interfering factors such as $Cl^-$, $SO_4^{2-}$, $NO_3^-$ and dissolved oxygen (DO) all, statistically, influenced the current response compared to the response from synthetic phosphate solutions at pH 8.00, DO of 8.54 mg $O_2$/L and the absence of interfering ions. Tests on real wastewater samples verified the effect of the interfering factors, as phosphate measurements from three different sampling points (influent, activated sludge mixed liquors and effluent) did not correlate favourably with measurements acquired from a specialised laboratory. The success of this sensor is probably dependent on the simultaneous measurement of, or the calibration for, interfering parameters. However, the former approach would most likely require additional probes to measure these interfering parameters and the latter would probably require a complex calibrating matrix to account for all the interfering parameters. Nonetheless, variations of such sensors reviewed in this paper and their encouraging results offer an optimistic field of improvement on the design of the sensor studied in this paper for it to be employed on real wastewater systems.

# ACKNOWLEDGEMENTS

First of all, I would like to thank God for all of his blessings and guidance in these definitive times in my life. I would also like to express my gratitude to Thames Water who sponsored this thesis project and made available all facilities imperative to its success. In addition, thanks are also due to Petroleum Development Oman for sponsoring my educational endeavour in the United Kingdom. I would also like to acknowledge the contributions of my academic advisor, Dr. Ana Soares, and my industry advisors, Dr. Ben Martin and Dr. Eve Germain-Cripps. I am also indebted to my family and friends who provided me with the love and care needed to deliver this project in its final form. Lastly, I would like to thank Cranfield University and all faculty members in the Water Science Institute who spared no effort in making my educational experience at Cranfield fulfilling and smooth.

# TABLE OF CONTENTS

# LIST OF FIGURES

# LIST OF TABLES

# LIST OF EQUATIONS

# LIST OF ABBREVIATIONS

| | |
|---|---|
| P | Phosphorus |
| WWTP | Wastewater Treatment Plant |
| UK | United Kingdom |
| DEFRA | The UK's Department for Environment, Food and Rural Affairs |
| MIP | Molecular Imprinted Polymers |
| LOD | Limit of Detection |
| UV-Vis | Ultraviolet-Visible |
| ICP-OES | Inductively Coupled Plasma Optical Emission Spectroscopy |
| ISE | Ion-Selective Electrodes |
| DO | Dissolved Oxygen |
| DI | Deionised |
| CV | Cyclic Voltammetry |
| ANOVA | Analysis of Variance |

# 1 Introduction

The control of phosphorus (P) levels in the discharges of wastewater treatment plants (WWTPs) has been a cause for concern for Water Utilities because of its adverse effects on the environment and rigorous regulations. The United Kingdom (UK) has adopted the Urban Waste Water Directive (*European Directive 91/271/EC*) that sets the phosphorus discharge consent at 1 mg/L or 2 mg/L, depending on the size of the treatment works (Council of the European Union, 1991). However, in accordance with the Water Framework Directive (*European Directive 2000/60/EC*), phosphorus consents are now granted on a case-by-case basis, depending on the receiving water body's quality (Council of the European Union, 2000). For instance, the phosphorus limit in one of the most environmentally vulnerable areas in the UK is as low as 0.0196 mg P/L; a limit set for the Axe catchment by the Environment Agency (Glavan, White and Holman, 2012). The aforementioned regulations limiting phosphorus discharge levels are mainly put in place to safeguard against eutrophication. Eutrophication is defined as the excessive growth of plant life due to nutrient (nitrogen and phosphorus) enrichment in water bodies. The existence of excess amounts of these nutrients aids in the uncontrolled growth of phytoplankton that in turn, deteriorates water bodies (Matsuo *et al.*, 2001). Therefore, to protect the environment from detrimental events, such as eutrophication, the Water Utilities require more intensive processes that enable the reduction and monitoring of phosphorus levels.

## 1.1 Methods for Phosphate Measurement in Wastewater

Currently, a molybdenum blue method described by Murphy and Riley (1962) is used to determine phosphorus levels in wastewater, which largely exists in the form of phosphates. This method, also known as the colorimetric method, utilises a solution of ammonium molybdate containing sulfuric acid, ascorbic acid and potassium antimonyl tartrate. The described solution readily reacts with phosphate ions ($PO_4^{2-}$) producing a blue-purple compound that obeys Beer's law (Murphy and Riley, 1962). The UK's Department for Environment, Food and Rural Affairs (DEFRA) (1992) recommends the colorimetric method as the standard method for the determination of phosphates, mainly due to its sensitivity (Udnan *et al.*, 2005). Although the colorimetric method is the predominant method used by the Water Utilities and the Environment Agency, the method suffers a range of drawbacks. For instance, as already described, the method utilises a mixed reagent that contains strong acids that are potentially toxic and limiting in terms of their application in online real time systems (Korostynska, Mason and Al-Shamma'a, 2012). Additionally, the mixed reagent does not keep for more than 24 hours. Therefore, fresh batches of the mixed reagent need to be prepared as required (Murphy and Riley, 1962). Moreover, the colorimetric method suffers from the interference of silicate and turbidity (Clesceri *et al.*, 1998). Hence, the drawbacks are labour intensive in nature and require the use of consumables, driving the need for a novel methodology to address them.

Consequently, numerous methods to detect phosphate have been proposed and studied. Reviews by Engblom (1998), Villalba *et al.* (2009) and Warwick, Guerreiro and Soares (2013) summarise these efforts. Although multiple techniques could potentially replace the current standard method, none have yet done so.

**1.1.1 Synthetic Molecular Recognition**

Synthetic molecular recognition techniques have been widely proposed to detect phosphate in wastewater. These techniques utilise the structural variety of anions to concoct complementary molecules to identify them. For example, to identify phosphates or sulphates, a receptor would be synthesised to recognise tetrahedral anions. Taking into consideration the size and degree of rigidity of the cavity used to bind with the analyte, these molecules usually show high affinity for specific target anions that allows for selective detection. One promising approach for the direct quantification of phosphate in wastewater monitoring applications is molecular imprinted polymers (MIPs) designed as phosphate receptors. Phosphate, as the target analyte, is bound to specific cavities formed during polymerisation and subsequent removal of target analyte templates (Warwick, Guerreiro and Soares, 2013). Thiourea based MIP receptors, combined with conductance based transducers, have produced phosphate sensors that are selective enough and have a level of detection appropriate for applications in the water industry (Warwick *et al.*, 2014). Further work by Warwick *et al.* (2014) on thiourea based MIP phosphate receptors conceived a membrane format design that has the ability to cause changes in conductance in the presence of phosphate compounds. The design has a limit of detection (LOD) of 0.16 mg P/L and a linear response range between 0.66 and 8 mg P/L. The sensor produced results in two minutes without the need for additional reagents. Such a design bears great potential due to the fact that it abates the main drawbacks with the standard colorimetric method. Although the LOD of Warwick *et al.* (2014)'s design is below the 1 and 2 mg P/L limit prescribed by the Urban Waste Water Directive, the Water Framework Directive demands lower detection limits. Nonetheless, while a lower LOD would have been preferable, the linear range achieved by Warwick *et al.* (2014) is opportune for further improvements.

Work presented by Soares *et al.* (2016) optimises and integrates the membrane design by Warwick *et al.* (2014) onto a probe. The new integrated sensor is composed of a MIP membrane mounted on a tip of a pH electrode. The novelty of this sensor rests on the MIP synthesised, which can hold cavities with the right shape to allow orthophosphates to fit in. Once phosphates occupy these cavities, changes in the potentiometric response of the sensor can be correlated with phosphate concentrations.

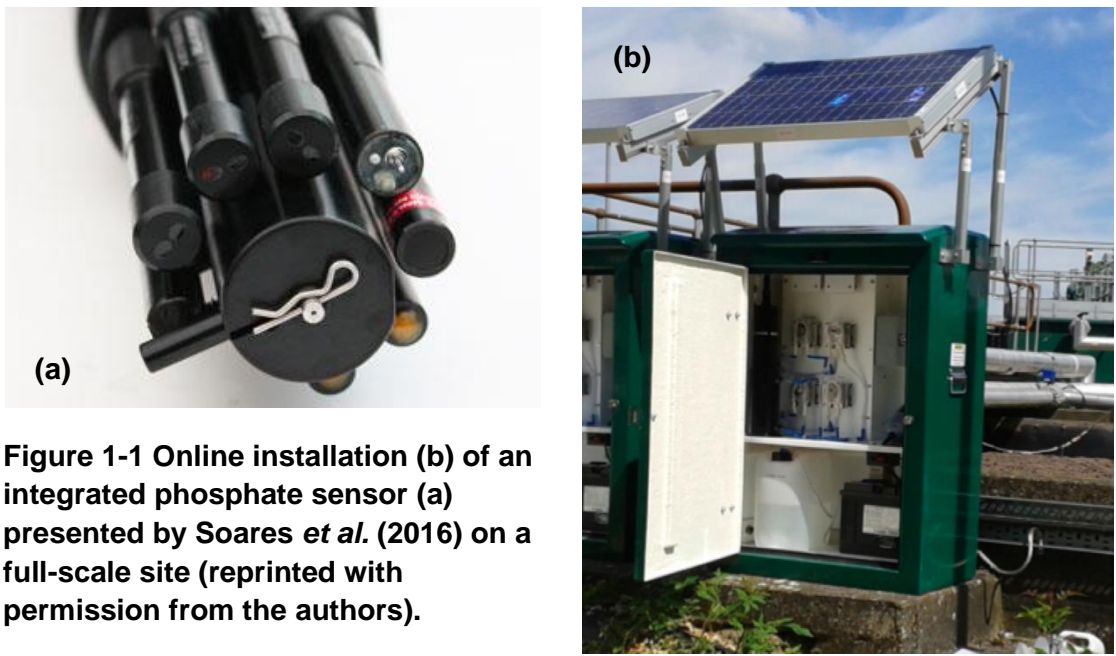

**Figure 1-1 Online installation (b) of an integrated phosphate sensor (a) presented by Soares *et al.* (2016) on a full-scale site (reprinted with permission from the authors).**

The integrated phosphate probe proposed by Soares *et al.* (2016) had been trialled on a full-scale site. The phosphate sensor was inserted onto an electrode housing with other sensors and was incorporated onto an online automated instrument, as Figure 1-1 illustrates. The sensor made nearly 5000 readings for over 3 months without the need for reagents, sample preparation or calibration. The readings were in agreement with phosphate concentrations acquired through colorimetric tests.

**1.1.2 Optical Detection**

Optical methods to detect phosphate are based on photo sensors that measure the wavelength of distinct colours. These colours can result from the addition of specific reagents that react with the target analyte (Korostynska, Mason and Al-Shamma'a, 2012). The molybdenum blue method described by Murphy and Riley (1962) represents an example of an optical method to detect phosphate.

**1.1.2.1 Colorimetric Methods**

Various colorimetric methods have been suggested and developed for the detection of phosphate in wastewater. However, most of these methods have their development and stability for long-term use hindered by interfering ions and fouling from solid particles. These disadvantages, in addition to inadequate detection limits for environmental use, aid in maintaining the standard molybdenum blue method as the dominant colorimetric method for phosphate detection (Warwick, Guerreiro and Soares, 2013). Hence, efforts to integrate the colorimetric method onto an online system have been made.

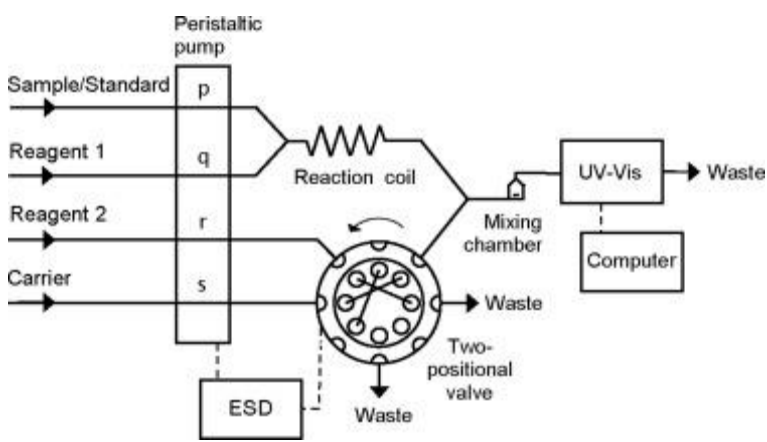

**Figure 1-2 A schematic of a colorimetric flow injection system for the simultaneous detection of phosphate and silicate proposed by Kozak *et al.* (2015) (reprinted with permission from Elsevier).**

Figure 1-2 depicts a colorimetric method to simultaneously determine phosphate and silicate ions. The method, proposed by Kozak *et al.* (2015), utilises ammonium molybdate to produce molybdophosphoric and molybdosilicic acids. Oxalic acid is then injected to decompose molybdophosphoric acid. The resulting reagent mixture is run through an ultraviolet-visible (UV-Vis) spectroscope to produce a peak in which the area of the peak correlates with the phosphate concentration and the absorbance registered at the peak's minimum correlate with the silicate concentration in a sample. Although the proposed method by Kozak *et al.* (2015) depends on chemical consumables, the online method produced results in about three minutes with detection limits of 0.054 mg P/l and 0.092 mg Si/L.

**1.1.2.2 Luminescence Methods**

Luminescence methods and colorimetric methods differ in terms of the interaction between light radiation and matter. Whereas colorimetric methods measure the absorbance of light by a specific analyte, luminescence methods measure the emission of light by a specific analyte in a sample. The two methods differ but relate in some capacity. Michalos and Rouff (2015) conducted a comparison between the standard colorimetric method and the Inductively Coupled Plasma Optical Emission Spectroscopy (ICP-OES), a luminescence method of analyte detection, on wastewater samples to detect phosphate. This study is important because various investigations had been carried out to evaluate these two techniques for phosphate detection on agricultural samples but few had been conducted on wastewater samples (Cantarero *et al.*, 2002 and Matula, 2010). The study demonstrated that both methods provide results with similar accuracy, with the ICP-OES method being advantageous in the sense that it requires less-to-no dilution of samples, depending on the source matrix containing the phosphate.

## 1.1.3 Electrochemical Sensing

Electrochemical methods have been gaining traction because of their sensitivity and selectivity in analyte detection in environmental and waste samples. Such methods are promising because of their compactness, rapid response and direct measurement of analyte in solution (Surkar and Karwankar, 2016).

### 1.1.3.1 Ion-Selective Electrodes

Ion-selective electrodes (ISE), which employ neutral and charged carriers infused in a polymeric membrane, have been investigated for phosphate measurement in wastewater samples (Bobacka, Ivaska and Lewenstam, 2008 and Crespo, 2017). ISEs are advantageous in terms of exhibiting low detection limits, overcoming reflective index errors and turbidity interference and eliminating the need for chemical consumables. However, they are disadvantageous in terms of selectivity, stability and precision (Kolliopoulos, Kampouris and Banks, 2015 and Crespo, 2017). Nonetheless, ISEs' abatement of the main drawbacks impeding the standard colorimetric method renders it suitable for future development to overcome its downsides.

### 1.1.3.2 Solid State Ion-Selective Electrodes

Solid state ion-selective electrodes differ from conventional ion-selective electrodes primarily in the way the two electrodes are constructed and maintained. Conventional ISEs contain liquid contact, referred to as inner filling solution. This solution is sensitive to evaporation and changes in sample temperature and pressure. Also, osmotic differences between samples and the inner filling solution can cause a net water transport through the ISEs' membrane that, in turn, can result in causing the inner filling solution's volume to change and the membrane's delamination (Lindner and Gyurcsanyi, 2009). Hence, conventional ISEs must be maintained and handled with care. With the rising need for robust and sustainable electrodes, it became optimal to phase out the inner filling liquid. Solid state ISEs do this by replacing the cumbersome fluid with a solid contact rendering an all-solid electrode design (Jinbo Hu and Buhlmann, 2016). Thus, solid state ISEs offer huge potential in providing an alternative electrochemical sensing device for phosphate detection. However, challenges with sensitivity to dissolved oxygen (DO) levels, interference by different ions, pretreatment and polishing necessities hinder the manufacturing of a reliable phosphate detector (Warwick, Guerreiro and Soares. 2013). Therefore, investment is still required to develop more robust solid state ISEs to replace the standard colorimetric method for phosphate detection.

## 1.2 Cobalt Based Electrochemical Detection of Phosphate

It had been postulated that Cobalt (Co) based ion-selective electrodes exploit a cobalt oxide (CoO) layer for the detection of phosphate in wastewater. A study conducted by Xiao *et al.* (1995) explains this affinity in terms of the guest-host chemistry of a nonstoichiometric compound: in this case, the compound being cobalt oxide. Xiao *et al.* (1995) suggest that the cobalt oxide layer reacts with DO in the sample forming cavities that could host phosphate compounds. The agglomeration of phosphate compounds in these cavities cause a potentiometric change that can be correlated with phosphate concentrations. Conversely, Meruva and Meyerhoff (1996) propose that the potentiometric change is actually caused by a mixed potential response prompted by the slow oxidation of cobalt, the simultaneous reduction of oxygen and $Co^{2+}$, and, in the presence of phosphate, the precipitation of cobalt phosphate on the electrode surface; Equations (1-1) - (1-3).

$$3 \cdot CoO + 2 \cdot H_2PO_4^- + 2 \cdot H^+ \leftrightarrow Co_3(PO_4)_2 + 3 \cdot H_2O \ (At\ pH\ 4.0) \qquad \textbf{(1-1)}$$

$$3 \cdot CoO + 2 \cdot HPO_4^{2-} + H_2O \leftrightarrow Co_3(PO_4)_2 + 4 \cdot OH^- \ (At\ pH\ 8.0) \qquad \textbf{(1-2)}$$

$$3 \cdot CoO + 2 \cdot PO_4^{3-} + 3 \cdot H_2O \leftrightarrow Co_3(PO_4)_2 + 6 \cdot OH^- \ (At\ pH\ 10.0) \qquad \textbf{(1-3)}$$

Regardless, both theories emphasise the susceptibility of the potentiometric response to DO levels, with the two theories highlighting the importance of the cobalt oxide layer in the detection mechanism.

Multiple cobalt based ISE designs have been recommended and investigated for phosphate detection. De Marco, Pejcic and Chen (1998) investigated a cobalt wire electrode in a carrier containing potassium hydrogenphthalate (KHP) at pH 5. The ISE showed a dynamic linear range between 3.10 - 310 mg P/L, which compares favourably with previous studies (Xiao *et al.*, 1995). The LOD was 0.0929 mg P/L; however, at low phosphate levels, the electrode showed low reproducibility. Additionally, the research team compared the performance of the electrode against a standard spectroscopic method with effluent samples from a WWTP. The research team concluded that the raw data the electrode provided had an error of 100% due to chloride ion ($Cl^-$) interference. A correction factor implemented to account for the error produced data to an accuracy of 5% relative. Nonetheless, the implementation of the correction factor requires simultaneous measurement of $Cl^-$ in the sample.

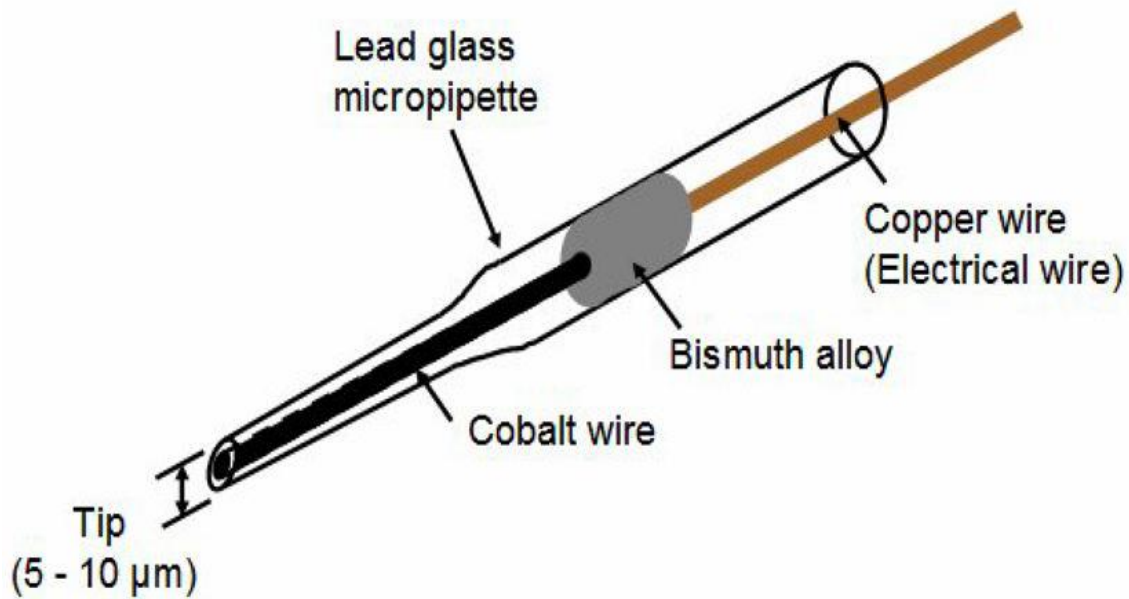

**Figure 1-3 Finished phosphate microelectrode with a tip diameter of 10 μm proposed by Lee, Seo and Bishop (2009) (reprinted with permission from Elsevier).**

Moreover, Lee, Seo and Bishop (2009) completed a comprehensive assessment of the performance of a cobalt-based microelectrode for the detection of phosphate in biological treatment processes and natural waters in correlation with eutrophication. The electrode's potentiometric response showed a linear correlation between phosphate concentrations of 0.310 - 3100 mg P/L with a LOD of 0.264 mg P/L. The electrode had a response time of 1 minute but alkalinity and DO levels significantly interfered with the electrode's readings.

The presence of anions such as carbonates ($CO_3^{2-}$) and hydroxides ($OH^-$) can cause reactions with $Co^{2+}$ to form $CoCO_3$ and $Co(OH)_2$ that affects the electrode's response. In terms of the DO levels, the research team concluded that DO interfered with the binding mechanism between cobalt oxide and phosphates which, in turn, resulted in an inversely proportional relationship between the potentiometric response and DO levels. This finding is consistent with the study done by De Marco, Pejcic and Chen (1998). Although the assessment conducted by Lee, Seo and Bishop (2009) on the phosphate microelectrode brings about encouraging findings, more field tests are required to substantiate the implementation of such an electrode in different types of wastewater.

Ding *et al.* (2015) examined a similarly designed cobalt based phosphate micro-sensor on lake waters and sediments. The electrode showed high sensitivity for phosphates with a LOD as low as 0.310 μg P/L in synthetic single ion solutions. The electrode's response varied from 5 seconds at high phosphate concentrations to 100 seconds at the lowest phosphate concentration detectable. Nitrates ($NO_3^-$), nitrites ($NO_2^-$) and sulphates ($SO_4^{2-}$), commonly found in freshwaters, were found to have statistically insignificant interference on the electrode's performance at low concentrations. However, at high $SO_4^{2-}$ concentrations, the electrode's performance was statistically impacted by interference. In addition, when the electrode's performance was tested against ion chromatography, it performed positively on environmental samples. However, as previously cited, DO levels interfered heavily with the electrode's readings. Additionally, environmental samples with high $SO_4^{2-}$ concentrations suppressed the electrode's performance, as concluded in the synthetic single ion samples experiments. Overall, microelectrodes demonstrate motivating results, both with synthetic single ion samples and wastewater samples after accounting for interfering species, which places the technology at a compelling status for further development.

Furthermore, Zhu, Zhou and Shi (2014) had studied electroplating cobalt on the surface of screen-printed electrodes as a potentiometric method for phosphate detection. The all solid-state ISE exhibited a phosphate-sensitive potential response between 0.310 - 3100 mg P/L, yielding a LOD of 0.0979 mg P/L in an acidic solution (pH of 4.00). With a relative standard deviation of 0.5%, these electrodes demonstrate high reproducibility with a response time of 40 seconds. Nevertheless, as with the previous three designs, DO levels significantly affected the electrodes response, corroborating the effect of oxygen on the interaction between cobalt oxide and phosphate. Also, comparing the electrode's performance with a standard spectroscopic method on real wastewater effluent samples produced results similar to those acquired by De Marco, Pejcic and Chen (1998). A correction factor for $Cl^-$ needed to be introduced to obtain results with an accuracy of 5%. Additionally, it was recommended that calibration curves for each pH of interest be established for more accurate results. In sum, with further investments placed in cobalt based screen-printed electrodes, the technology can progress to an advantageous position to replace the standard colorimetric method.

Table 1-1 summarises the cobalt based electrochemical sensors for phosphate detection. Although the mechanism behind these electrodes' response is a subject of debate, the linear range and the LOD of these electrodes in synthetic single ion solutions presents some alluring possible replacements for the standard colorimetric method. However, measures to circumvent the issues that hinder the implementation of these electrodes need to be taken to enact such replacements.

**Table 1-1 Summary of cobalt based electrochemical sensors for phosphate detection.**

| Method | Linear Range (mg P/L) | LOD (mg P/L) | Major Issues | Reference |
|---|---|---|---|---|
| Surface-modified cobalt based sensor | 0.310 - 310 | 0.155 | Sensitive to DO levels | (Xiao *et al.*, 1995) |
| Cobalt electrode | 3.10 - 310 | - | Not tested on wastewater samples | (Meruva and Meyerhoff, 1996) |
| Cobalt wire ISE | 3.10 - 310 | 0.0929 | $Cl^-$ interference | (De Marco, Pejcic and Chen, 1998) |
| Cobalt based microelectrode | 0.310 - 3100 | 0.264 | Alkalinity interference | (Lee, Seo and Bishop, 2009) |
| Cobalt based microsensor | - | $3.10 \times 10^{-3}$ | Sensitive to High $SO_4^{2-}$ Levels | (Ding *et al.*, 2015) |
| Cobalt based screen-printed electrode | 0.310 - 3100 | 0.0979 | Response specific to pH | (Zhu, Zhou and Shi, 2014) |

Table 1-2 highlights the desired key attributes in a functional wastewater phosphate sensor. A rapid response at low detection limits with minimal interference from foreign ions is vital for the successful integration of sensors in real time wastewater streams. Moreover, reproducible and accurate readings are of importance, especially with the introduction of stringent limits with the Water Framework Directive. Additionally, sensors' compactness, cost and independence from chemical need dictate the sensors' potential in having a profound impact on the efficiency of phosphate detection in the wastewater industry.

**Table 1-2 Key characteristics desired for an ideal phosphate sensor.**

| Sensor Characteristics | Desired Attribute |
|---|---|
| Response Time | 2 – 4 minutes |
| Interference | No to minimal interference from competing ions |
| LOD | < 0.100 mg P/L |
| Resilience | Precise and accurate measurements even at extreme wastewater parameters |
| Reliability | • Precise and accurate measurements when sensor is compared with standard methods<br>• Precise and accurate measurements when sensor is tested on real wastewater samples |
| Maintenance | • No to minimal calibration requirements<br>• No consumables required (Chemicals)<br>• Not impaired by solids and suspended matter |
| Cost | Minimal |

Specifically, this study tested a screen-printed cobalt based sensor for real time measurement of phosphate in municipal wastewater treatment systems. The sensor's reproducibility and resilience was tracked using a single ion model solution. Furthermore, as noticed in the literature review of similar designs (Table 1-1), the success of the technique is probably dependent on the simultaneous measurement of interfering parameters. Therefore, interference from DO, $Cl^-$, $SO_4^{2-}$ and $NO_3^-$ was assessed. Finally, field tests at an operational wastewater treatment works was carried out to evaluate the sensor's applicability in real wastewater samples.

# 2 Materials and Methods

## 2.1 Materials

Chemical reagents potassium dihydrogen phosphate ($KH_2PO_4$), sodium chloride (NaCl), and acetic acid ($CH_3COOH$) were purchased from Acros Organics (Belgium). Sodium sulphates ($Na_2SO_4$), potassium nitrate ($KNO_3$) and sodium hydroxide (NaOH) were supplied by Fisher Scientific (UK). Nitrogen gas ($N_2$), used for deoxygenation, was acquired from BOC Limited (UK). The screen-printed cobalt based phosphate sensors examined were printed on a ceramic substrate and consisted of a working electrode made of carbon conductive ink doped with cobalt-phthalocyanine (4mm in diameter), an auxiliary electrode made of non-modified carbon and a pseudo-reference electrode made of silver (Rama *et al.*, 2012). The sensors' potentiometric response was assessed using a bipotentiostat provided by DropSens (Spain).

## 2.2 Assessment of the Potentiometric Response of the Sensors

### 2.2.1 Effect of pH

Synthetic single ion model solutions with different phosphate levels (0, 0.100, 1.00, 10.0, 25.0, 50.0 and 100 mg P/L) were prepared at different pH (4.00 and 8.00) to determine the effect of pH on the sensor's potentiometric response. A pH probe was provided by Hach (UK). The phosphate solutions were prepared by first making a bulk 100 mg P/L solution using $KH_2PO_4$. Phosphate solutions at different concentrations were then prepared by diluting the bulk phosphate solution with deionised (DI) water. Finally, the pH of the final solutions was manipulated using $CH_3COOH$ and NaOH to obtain the desired pH.

The cobalt based sensors were preconditioned in DI water for 15 minutes so that a CoO layer formed (Meruva and Meyerhoff, 1996). The sensors were then immersed in samples for 10 minutes before commencing with cyclic voltammetry (CV). DropView, a software program provided with the bipotentiostat, was used to record the voltammograms at a scan rate of 50 mV/s from -1.4 V to 1.4 V. Blank samples of DI water and phosphate single ion synthetic solutions were trialled initially to identify the peak representing the cobalt-phosphate interaction. The intensity of the peak provided by the cyclic voltammograms was recorded in terms of their current response and correlated with their respective phosphate concentration.

### 2.2.2 Effect of Interfering Ions

Interference from $Cl^-$, $SO_4^{2-}$ and $NO_3^-$ (100 mg/L each) was assessed by dissolving NaCl, $Na_2SO_4$ and $KNO_3$ at different phosphate levels at a pH of 8.00. The phosphate concentrations in solution were obtained as detailed above. However, after the different phosphate solutions were prepared, appropriate amounts of the interfering ions were added before correcting the pH to 8.00. The sensors' response in solution was then evaluated, as described above, to observe the changes in the peak currents' intensity.

### 2.2.3 Effect of Dissolved Oxygen

Interference from DO was analysed by purging $N_2$ at different phosphate levels at a pH of 8.00. DO reduction procedure by $N_2$ purging elaborated by Butler, Schoonen and Rickard (1994) was followed to reduce the DO level to 1.00 mg $O_2$/L. A DO probe was provided by Hach (UK). The pH of the solutions was adjusted to 8.00 prior to the deoxygenation step to avoid ambient oxygen dissolving again into the solutions while adding reagents.

### 2.2.4 Tests on Real Wastewater Samples

To ensure the reliability of the acquired results, the sensors were tested on real wastewater samples. Samples from Oxford Sewage Treatment Works that serves a population equivalent of 216,000, operated by Thames Water, were collected from three sample points; the work's influent, activated sludge mixed liquors and effluent. The samples were left to settle for a day before phosphate measurements carried out using the sensors took place. Additionally, the samples' were also sent to a Thames Water analysis laboratory for characterisation to validity the sensors' readings.

# 3 Results

## 3.1 Phosphate Peak Identification

To begin with, experiments were first conducted to identify current peaks characteristic to phosphate. Figure 3-1 depicts the cyclic voltammograms produced from the experiments. A clear amplification in current response and a shift in the peak potential were observed. The potentials at which the current peaked were in between -0.500 V - -1.10 V depending on the phosphate content.

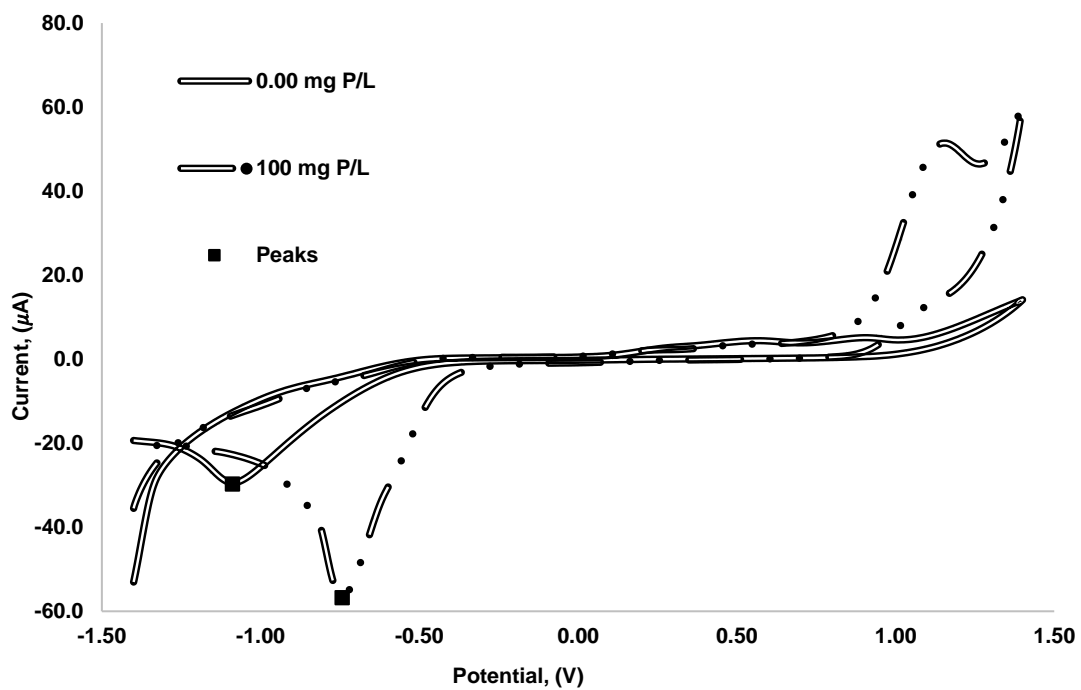

**Figure 3-1 Cyclic voltammograms produced from immersing the Cobalt based sensors in DI water with no phosphate (0.00 mg P/L) and in a solution of 100 mg P/L; all at pH 8.00.**

## 3.2 Effect of pH

Figure 3-2 illustrates the calibration curves between the intensity of the peak currents and their respective phosphate concentration at pH 4.00 and 8.00. As the figure shows, increasing phosphate concentrations yield higher intensity peak currents. This proportionality between the peak currents and the respective phosphate concentration was linear with $R^2$ values of 0.981 and 0.972 at pH 8.00 and 4.00, respectively.

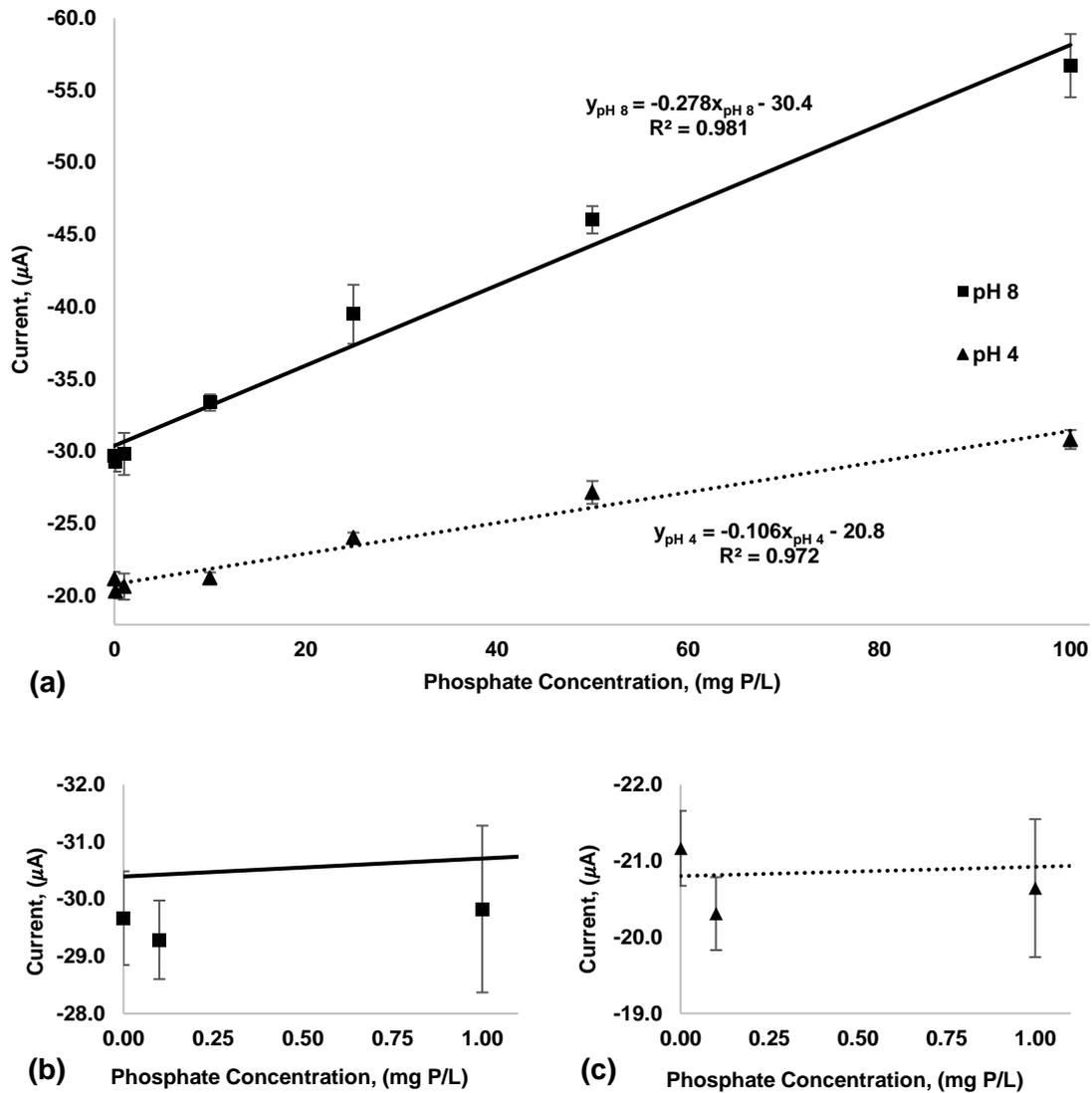

Figure 3-2 Peak current and phosphate concentration's calibration curves at pH of 4.00 and 8.00 (a). Additionally, a detailed representation of the calibration curves at the lower end of phosphate concentrations at pH 8.00 (b) and 4.00 (c). The error bars show the standard deviation of 4 replicates.

Furthermore, the sensor exhibited an affinity for phosphates with a LOD of (18 ± 10) mg P/L at pH 4.00 and (6 ± 8) mg P/L at pH 8.00. The errors associated with the LOD were propagated from the current response and the calibration curves (University of California-Davis, 2017). Also, although the sensors had to be pre-treated for 25 minutes, they generated a response in just under 2 minutes after the CV test was initiated. An analysis of variance (ANOVA) was conducted, which suggested that there was a significant difference between the current response provided at different pH, statistically (P<0.05).

Figure 3-3 shows the calibration curves between the potentials at which the peak currents occur and the logarithmic phosphate concentration. The figure depicts a directly proportional relation between the intensity of the peak potentials and the logarithmic phosphate concentrations. The $R^2$ values for the calibration curves at pH 4.00 and 8.00 are 0.942 and 0.960, respectively. Furthermore, as observed in Figure 3-3, the calibration curves are more linear at the higher end of the phosphate concentrations.

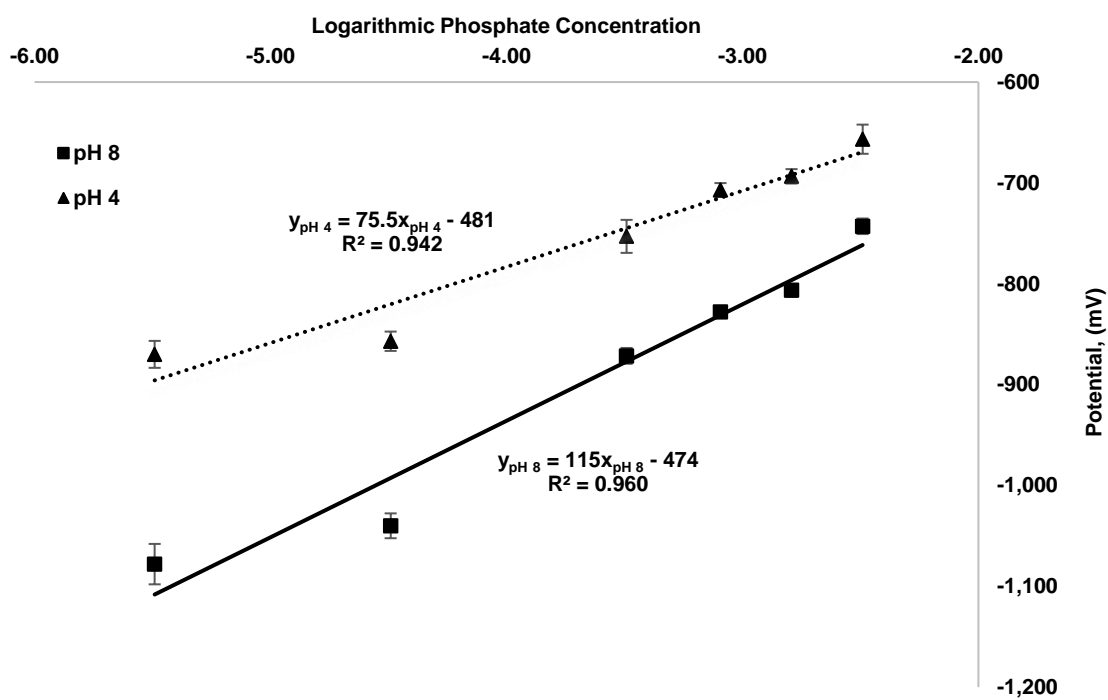

**Figure 3-3 Peak potential and phosphate concentration's calibration curves at pH of 4.00 and 8.00. The error bars show the standard deviation of 4 replicates.**

## 3.3 Effect of Interfering Ions

Figure 3-4 depicts the calibration curves of the phosphate solutions with different interfering ions ($Cl^-$, $SO_4^{2-}$ and $NO_3^-$) compared to the calibration curve of synthetic phosphate solution at pH 8.00. The proportionality between the current response and phosphate concentration observed in Figure 3-2 was also observed in Figure 3-4. The $R^2$ values of the three calibration curves were 0.974, 0.982 and 0.966 for $Cl^-$, $SO_4^{2-}$ and $NO_3^-$'s curves, respectively.

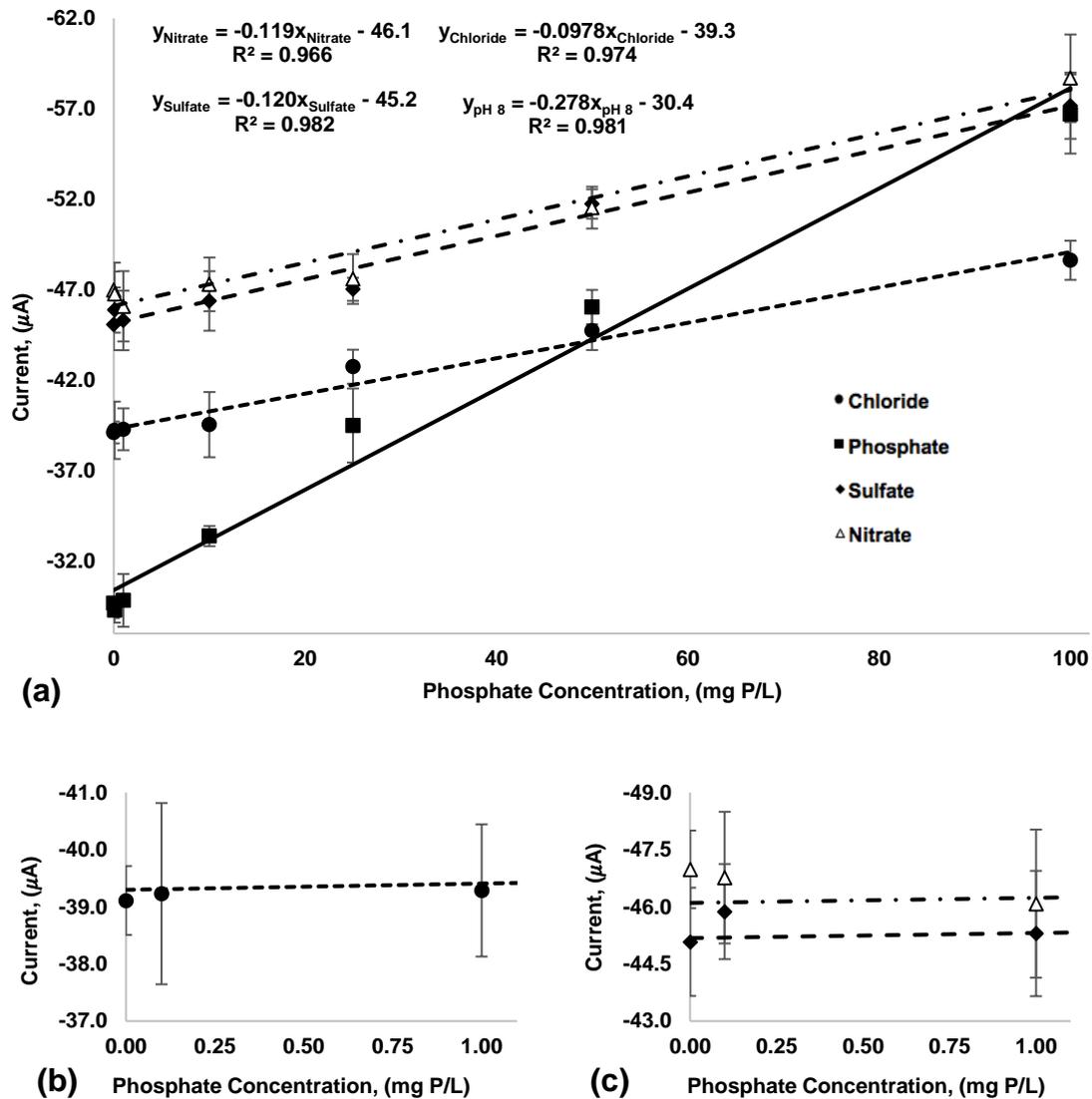

(a)
(b)
(c)

**Figure 3-4 Peak current and phosphate concentration's calibration curves at pH 8.00 with different interfering ions (a). Additionally, a detailed representation of the calibration curves at the lower end of phosphate concentrations at pH 8.00 for chloride (b) and sulphate and nitrate (c). The error bars show the standard deviation of 4 replicates.**

ANOVA statistical analysis tests conducted on the different calibration curves shown in Figure 3-4 resulted in suggesting that there were significant differences in the current responses between the phosphate solutions that contained interfering ions and the synthetic phosphate solutions that contained no interfering ion ($P<0.05$).

## 3.4 Effect of Dissolved Oxygen

Figure 3-5 illustrates the calibration curve of phosphate solutions with 1.00 mg $O_2$/L compared to the calibration curve of synthetic phosphate solutions at pH 8.00 which had a DO level of 8.54 mg $O_2$/L. The proportionality between the current response and phosphate concentrations observed was concurrent with the behaviour noted in Figure 3-2 and Figure 3-4. An ANOVA test was conducted that suggested that there was a significant difference between the current responses provided at different DO levels (P<0.05).

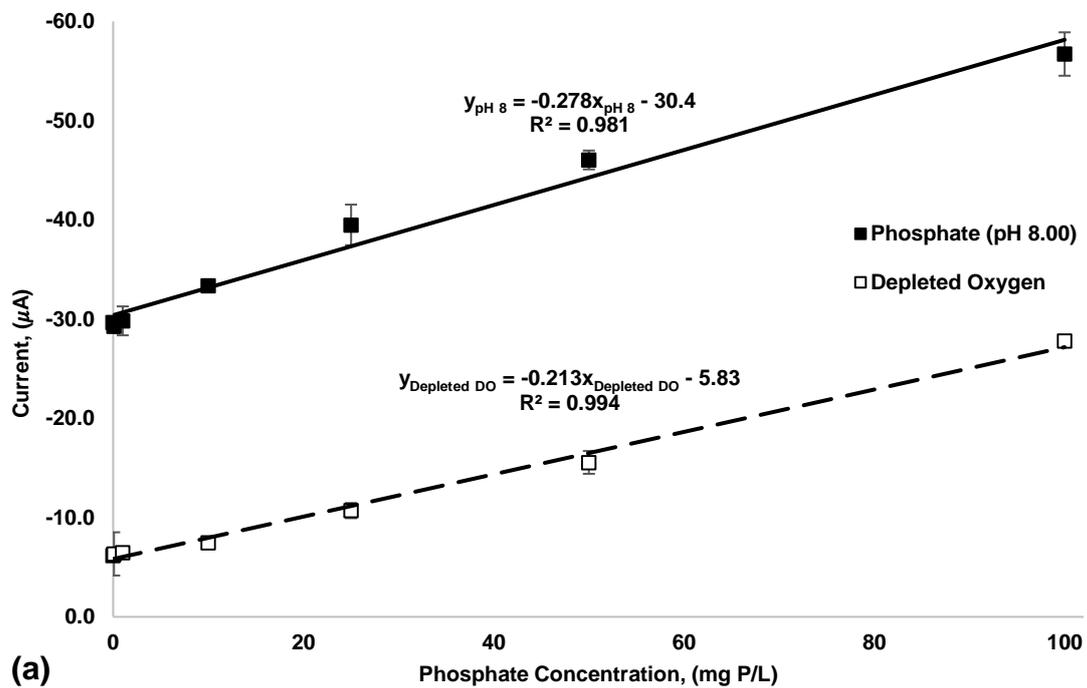

(a)

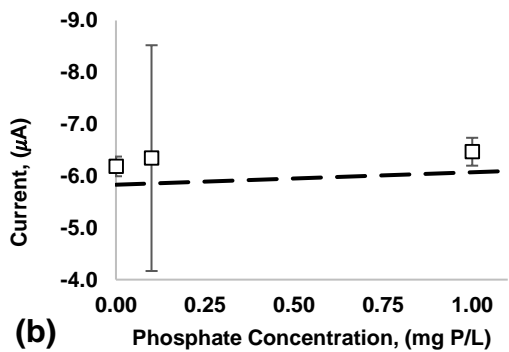

(b)

**Figure 3-5 Peak current and phosphate concentration's calibration curve at pH 8.00 with 1.00 mg $O_2$/L (a). Additionally, a detailed representation of the calibration curve at the lower end of phosphate concentrations (b). The error bars show the standard deviation of 2 replicates.**

## 3.5 Tests on Real Wastewater Samples

Figure 3-6 entails cyclic voltammograms tested on real wastewater samples from three sampling points; influent, activated sludge mixed liquors and effluent. Peaks between -0.500 - -1.00 V had been identified as the phosphate peaks in accordance with the experiments conducted on synthetic phosphate solutions. Therefore, the phosphate peak intensities for the influent, mixed liquors and effluent were −(4 ± 3) µA, −(33 ± 2) µA and −(46 ± 1) µA, respectively. Using the calibration equation between the current response and phosphate concentrations acquired at pH 8.00 with no interfering ions, the phosphate concentration corresponding to the peak current recorded for each wastewater sample was calculated. The peak current obtained from the influent sample was out of the range in which the calibration curve was constructed. Thus, no reading was attained from the influent sample but the readings for the mixed liquors and effluent samples were (11 ± 9) mg P/L and (58 ± 12) mg P/L.

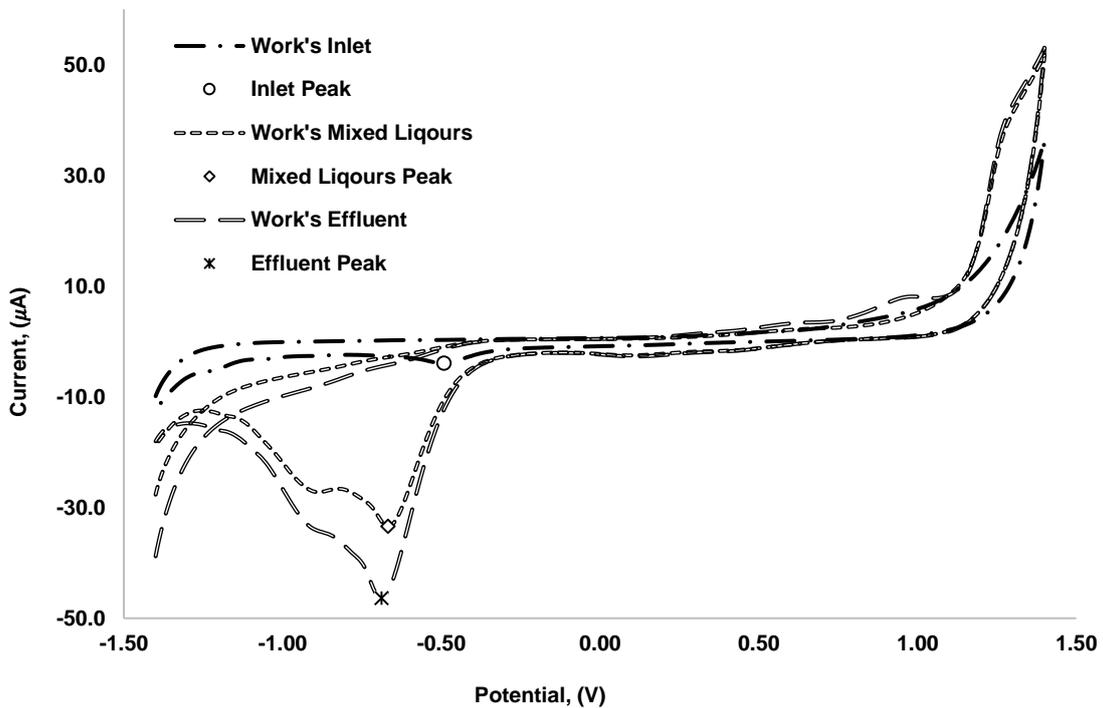

**Figure 3-6 Cyclic voltammograms of real wastewater samples from different sampling points.**

Table 3-1 summarises the analysis results received from Thames Water's analysis laboratory. Although the mixed liquors' phosphate concentration value was quite different between the measurement taken by the sensor and the characterisation's results, according to a two-tailed t-test, there was no significant difference between the two values, statistically. On the other hand, the effluent readings were significantly different.

**Table 3-1 Wastewater analysis results and the sensors' phosphate readings.**

| Component Name | Results Received | | | Units |
| --- | --- | --- | --- | --- |
|  | Influent | Mixed Liquors | Effluent |  |
| Temperature | 23.8 | 22.7 | 22.5 | °C |
| pH | 7.60 | 7.00 | 7.10 | - |
| DO | 0.900 | 5.99 | 8.50 | mg $O_2$/L |
| Phosphate | 3.87 | 0.210 | 0.160 | mg P/L |
| $Cl^-$ | 97.5 | 89.2 | 91.0 | mg $Cl^-$/L |
| $SO_4^{2-}$ | 73.4 | 125 | 124 | mg $SO_4^{2-}$/L |
| $NO_3^-$ | 0.885 | 87.2 | 105 | mg $NO_3^-$/L |
| Cobalt based sensor phosphate reading | Out of Range | 11 | 58 | mg P/L |

# 4 Discussion

## 4.1 Phosphate Peak Identification

Figure 3-1 illustrates the voltammetric response of the cobalt based sensors in DI water at pH 8.00 compare to a solution with a 100 mg P/L at pH 8.00 also. The pure DI water samples shape a peak at around -1.10 V that was previously reported to relate to the formation of the CoO layer; Equations (4-1) – (4-3) (Meruva and Meyerhoff, 1996).

$$2 \cdot Co + 2 \cdot H_2O \leftrightarrow 2 \cdot CoO + 4 \cdot H^+ + 4 \cdot e^-  \quad (4\text{-}1)$$

$$O_2 + 4 \cdot H^+ + 4 \cdot e^- \leftrightarrow 2 \cdot H_2O \quad (4\text{-}2)$$

$$2 \cdot Co + O_2 \leftrightarrow 2 \cdot CoO \quad (4\text{-}3)$$

The solution with a 100 mg P/L exhibited a more intense current peak with a shift in potential. This deviation was also justified by Meruva and Meyerhoff (1996), Equations (1-1) – (1-3), and further developed by De Marco, Pejcic and Chen (1998); Equations (4-4) – (4-6).

$$Co + H_2O \leftrightarrow CoO + 2 \cdot H^+ + 2 \cdot e^- \quad (4\text{-}4)$$

$$CoO + 2 \cdot KH_2PO_4 \leftrightarrow K_2[Co(HPO_4)_2] + H_2O \quad (4\text{-}5)$$

$$Co + 2 \cdot KH_2PO_4 \leftrightarrow K_2[Co(HPO_4)_2] + 2 \cdot H^+ + 2 \cdot e^- \quad (4\text{-}6)$$

The latter developed Meruva and Meyerhoff (1996) justification by elucidated that the enhancement in the charge-transfer process did not just depend on the phosphate content but also on the pH ( $[H^+]$ ).

## 4.2 Effect of pH

Although a linear calibration curve was successfully constructed at both pH (4.00 and 8.00), the cobalt sensors response was significantly different. In solution, phosphate can exist in different forms, as Figure 4-1 demonstrates. Furthermore, as Meruva and Meyerhoff (1996) postulated, in the presence of any form of phosphate in solution, cobalt phosphate can precipitate as Equations (1-1) - (1-3) detail. Therefore, this explains why the sensors are responsive at different pH values. However, the sensors' response was better at pH 8.00 than at pH 4.00 because, as De Marco, Pejcic and Chen (1998) outlined, diminishing hydron ion (H$^+$) concentrations enhanced the potential response of cobalt based phosphate sensors. This behavior has also been observed by Lee, Seo and Bishop (2009). Hence, the calibration curves and their corresponding LOD values for synthetic phosphate solutions at elevated pH (pH 8.00) were better than the solutions at pH 4.00.

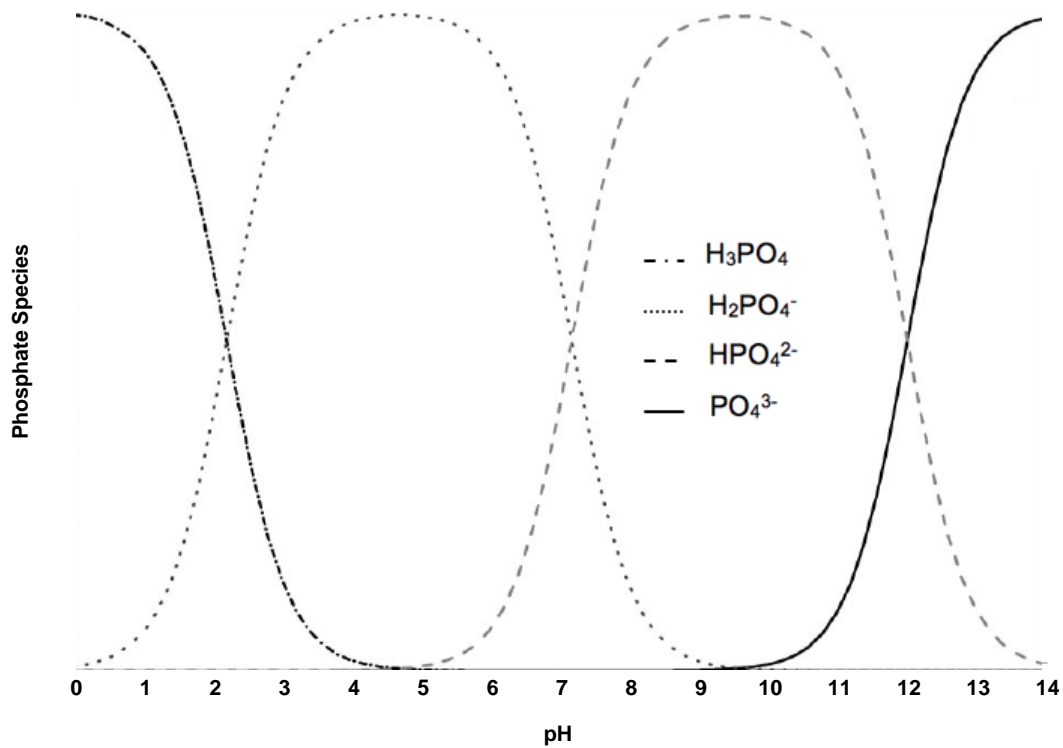

**Figure 4-1 Phosphate speciation diagram as a function of pH (the figure was adopted in part with permission from Eliaz and Sridhar. Copyright (2008) American Chemical Society).**

Although this study concentrates on the current response, Figure 3-3 provides another aspect that can be exploited to measure phosphate levels in wastewater samples through measuring the potential response of the sensors. The peak potential-logarithmic phosphate concentration calibration curves followed a linear response in agreement with the Nernst equation, which was previously suggested to explain the potential response of cobalt based phosphate sensors; Equation (4-7) (De Marco, Pejcic and Chen, 1998).

$$E = E^o - \frac{2.303 \cdot R \cdot T}{F} \cdot \log[P] \tag{4-7}$$

Where,

$E$: Sensor's potential response in V.

$E^o$: Potential response in V at specific conditions of pH.

$R$: The universal gas constant; 8.31 J/K·mol.

$T$: Temperature in K.

$F$: The Faraday constant; 96500 C/mol.

$[P]$: Phosphate concentration in M.

Nonetheless, De Marco, Pejcic and Chen (1998) and Xiao *et al.* (1995) all observed that the potential response followed a non-Nernst linearity at low logarithmic phosphate concentrations ($1 \times 10^{-4}$ - $1 \times 10^{-5}$). The same response was observed in Figure 3-3 at low logarithmic phosphate concentrations ($1 \times 10^{-4}$ - $1 \times 10^{-6}$), attributed to low solubility of cobalt and adsorbed phosphate species on the sensor's surface.

The LOD at pH 8.00 and 4.00 had relatively high errors associated at 8 and 10 mg P/L, respectively. This indicates the high instability of the current response used to calculate the LOD, which, in turn, questions the reproducibility and reliability of the sensors used. However, the high margin of error regarding the detection limits can also be attributed to the fact that the data points used to construct the calibration curves were dispersed in a trend around the curves. The way in which points are scattered around a proposed calibration curve provides useful information to validate curves because data points should be randomly scattered to deem any curve reliable (Van Loco *et al.*, 2002). A possible explanation of why the data points in the pH experiments were scattered around the calibration curve in a trend could be the instabilities in the pH, as, for some samples, the pH before the CV tests differed slightly from the pH following the tests. In sum, the results acquired from assessing the effect of pH on the sensor's performance suggest the dependency of the sensor's response on pH. Thus, the sensors need to be calibrated for different pH values.

## 4.3 Effect of Interfering Ions

An indirect method was employed to assess the effect of interfering ions: the interfering ion was kept constant, while the phosphate concentration was varied. Although linear calibration curves with phosphate were successfully constructed, statistically, all the interfering ions assessed influenced the sensors' response. Based on the gradient of each calibration curve, the best response being the steepest, $Cl^-$ influenced the sensors' response the most; A slope of -0.0978 for chloride compare to -0.120 for sulphate and -0.119 for nitrate. This observation is in agreement with findings by De Marco, Pejcic and Chen (1998), Lee, Seo and Bishop (2009) and Xiao *et al.* (1995). The interference from $SO_4^{2-}$ was statistically significant, too, but to a lesser extent. In a study carried out by Ding *et al.* (2015) on the influence of $SO_4^{2-}$ and $NO_3^-$ on cobalt based sensors, the authors recommended not using the sensors in high $SO_4^{2-}$ concentrations without re-calibrating. Nitrate had a very similar but statistically different response (P=0.04) to the response from $SO_4^{2-}$, which is consistent with findings produced by Ding *et al.* (2015). Overall, $Cl^-$, $SO_4^{2-}$ and $NO_3^-$ affected the sensors' responses in different magnitudes; $Cl^-$ having the highest influence and $SO_4^{2-}$ and $NO_3^-$ the least. Lee, Seo and Bishop (2009) and Xiao *et al.* (1995) reported the same trend.

Although the interfering ions impacted the sensors' response, compared to the calibration curve generated in the absence of these species, all the interfering ions maintained a good linear calibration response; substantiated by the $R^2$ values of each. This feature situates such cobalt based electrodes in an advantageous position against other phosphate measuring tools. Efforts to decrease the influence of interfering ions on cobalt based sensors had been reported in previous studies with particular success in lessening the effect of $Cl^-$. De Marco, Pejcic and Chen (1998) applied the Nicolsky equation in calculating a correction factor to account for the effect that the $Cl^-$ ion had on the sensors' response. Compared to phosphate measurements obtained by a standard spectroscopic method, the corrected response offered good agreement, signifying the reliability of such an approach. However, such a method would than require the simultaneous measurement of Cl$^-$ ions in sample, which limits the implementation of online measurements because of the unavailability of online tools to measure Cl$^-$ ions.

Altogether, although the slopes of the interfering ions' calibration curves were statistically different, they were close in value. This aspect can be exploited to formulate an approach to calibrate the sensors used in this study. However, as previously mentioned, simultaneous measurement of interfering ions might be an obstacle that would need to be circumvented.

## 4.4 Effect of Dissolved Oxygen

The effect of DO levels on the cobalt based sensors was also indirectly assessed. The DO concentration was reduced from 8.54 to 1.00 mg $O_2$/L through $N_2$ purging. A linear calibration curve with a better $R^2$ value was constructed at depleted oxygen levels with the calibration curves at the two DO levels being statistically different. However, the data points used to construct the calibration curves were not evenly distributed around the proposed curve, which could be attributed to the instability of the depleted oxygen samples. The DO levels in the samples tested for the effect of oxygen changed with time, prompting replicating the experiments only twice (rather than the four times conducted in all the other experiments). The effect of DO on the workings of cobalt based sensors for phosphate quantification had been reported in all papers concerned with the topic, including all those reported in Table 1-1. It is understood that the binding mechanism between CoO and phosphate is adversely affected at elevated oxygen levels (Lee, Seo and Bishop, 2009). This explains why the sensors' response at low oxygen levels, although shifted towards lower peak current values, had a slightly better $R^2$ value.

It is recommended to calibrate the cobalt based phosphate sensors at different DO levels. Although this approach would require simultaneous DO measurement, probes to do so are commercially available. Findings from Ding *et al.* (2015) led to analogous recommendations that were tested by the research's authors. The researchers compared phosphate measurements obtained by ion chromatography and cobalt based phosphate sensors. The measurements correlated accurately, especially after correcting for DO interference. Thus, such an approach would be efficient in suppressing the influence of DO.

## 4.5 Tests on Real Wastewater Samples

The calibration curve between the peak currents and the phosphate concentrations at pH 8.00 was used to ensure the reliability and validity of the cobalt based phosphate sensors' response. As Figure 3-6 and Table 3-1 demonstrate, the sensors' peak currents responses increased alongside rising DO levels with each wastewater sample. This behaviour is evident when comparing the cyclic voltammograms generated from the different wastewater samples and the voltammogram generated from the synthetic solutions. The cyclic voltammograms of the wastewater samples deviated from the synthetic solution's behaviour as the DO levels were different. This validates the high dependency of the cobalt based phosphate sensor on oxygen.

Although the DO levels between the effluent sample and the synthetic solution were alike, the phosphate sensors did not accurately report the phosphate level. This discrepancy could be credited to the high $Cl^-$ concentrations, as Table 3-1 details. The experiments conducted in this study and the reviewed literature each support the above finding. Additionally, the high $SO_4^{2-}$ and $NO_3^-$ concentrations in the wastewater samples could have added to the inaccuracies the cobalt based phosphate sensor experienced. One other possible source of error is the different pH values between the wastewater samples and the synthetic phosphate solutions. Figure 4-1 illustrates the sensitivity of the phosphate speciation in solution to pH values; the ratio between $H_2PO_4^-$ and $HPO_4^-$ change significantly between pH of 6.00 – 8.00. This fact could scramble the current response, as the formation of cobalt phosphate from $H_2PO_4^-$ (Equation 1-1), which occurs at low pH, is less effective than the formation of cobalt phosphate from $HPO_4^{2-}$ (Equation 1-2) (De Marco, Pejcic and Chen, 1998). Actually, Lee, Seo and Bishop (2009) tested cobalt based phosphate sensors at pH between 6.00 and 8.00. The study concluded that the sensors performed differently in this tight pH window, spotlighting the necessity of calibrating cobalt based phosphate sensors at pH in which the sensors are going to be used. The combined effect that the DO, interfering ions and pH had on the sensors response has evidently affected the measurements that were inaccurate and imprecise.

Methods of error propagation that utilise the standard deviations of the cobalt based phosphate sensors were applied to assess the precision of the sensors. The errors associated with the measurements acquired by the sensors were too large for the sensors to be deemed reliable. For instance, a two-tailed t-test, conducted to assess if the phosphate measurements provided by the laboratory and the sensor were similar, revealed that the two measurements were statistically alike although the two values were vastly different (11 $\pm$ 9 and .0.210 mg P/L). This false positive result could be traced to the imprecise results the sensors provided, rendered in the large standard deviation.

Table 4-1 characterises the cobalt based phosphate sensor tested in this study. The sensor's reproducibility and resilience is in question due to the heavy interference from different factors and the large errors associated with the sensors' readings. The success of this sensor is probably dependent on the simultaneous measurement of, or the calibration for, interfering parameters. However, the former approach would most likely require additional probes to measure these interfering parameters and the latter would probably require a complex calibrating matrix to account for all the interfering parameters. Nonetheless, variations of such sensors reviewed in Table 1-1, and their encouraging results, offer an optimistic field of improvement on the design of the sensor studied in this paper for it to be employed on real wastewater systems.

**Table 4-1 Key characteristics of the studied cobalt based phosphate sensor.**

| Sensor Characteristics | Desired Attribute |
|---|---|
| Response Time | < 2 minutes |
| Interference | Response interfered by DO levels and interfering ions in the below order:<br>$Cl^- > SO_4^{2-}$ & $NO_3^-$ |
| LOD | (6 ± 8) mg P/L at pH 8.00 and DO 8.54 mg $O_2$/L |
| Resilience | The sensors would need to be calibrated at different pH and DO levels. |
| Reliability | - Poor correlation when compared with standard methods<br>- Imprecise and inaccurate measurements when sensor is tested on real wastewater samples, due to the interfering factors mentioned above |
| Maintenance | - Sensor requires multiple calibrations<br>- No consumables required (Chemicals) |

# 5 Conclusions

In conclusion, in this study, the performance of a screen-printed cobalt based phosphate sensor was calibrated, tested for ion and DO interference, and, finally, tested on real wastewater samples. The sensors' current response correlated lineally with increasing phosphate concentrations between 6.00 – 100 mg P/L. A calibration curve established with synthetic phosphate solutions at pH 8.00 offered a LOD of (6 ± 8) mg P/L. pH, DO and interfering ions, such as $Cl^-$, $SO_4^{2-}$, $NO_3^-$, all statistically influenced the sensor's response. Test on real wastewater samples verified the effect of the interfering factors mentioned above, as phosphate measurements from three different sampling points (influent, activated sludge mixed liquors and effluent) did not correlate favourably with measurements acquired from a specialised laboratory.

# APPENDICES

# Appendix A – Effect of pH Data Analysis

Appendix A provides the raw data acquired from experiments conducted on synthetic phosphate solutions at different concentrations and pH accompanied with the cyclic voltammograms produced. The appendix also details all the data analysis carried out on the data set.

## A.1 pH 4.00

Table A-1 and Figure A-1 provide the data points observed in the cyclic voltammograms tested at different phosphate concentrations. For each concentration, 4 data replicates were taken. The average of the 4 readings was then calculated.

**Table A-1. Average raw data acquired from experiments conducted on synthetic phosphate solutions at pH 4.00.**

| Phosphorus Concentration (mg P/L) | Average Peak Current (uA) | Standard Deviation | Average Peak Potential (V) | Standard Deviation |
|---|---|---|---|---|
| 0.00 | -21.2 | 0.5 | -0.876 | 0.007 |
| 0.100 | -20.3 | 0.5 | -0.87 | 0.01 |
| 1.00 | -20.6 | 0.9 | -0.86 | 0.01 |
| 10.0 | -21.2 | 0.4 | -0.75 | 0.02 |
| 25.0 | -24.0 | 0.3 | -0.707 | 0.007 |
| 50.0 | -27.2 | 0.8 | -0.694 | 0.007 |
| 100 | -30.8 | 0.7 | -0.66 | 0.01 |

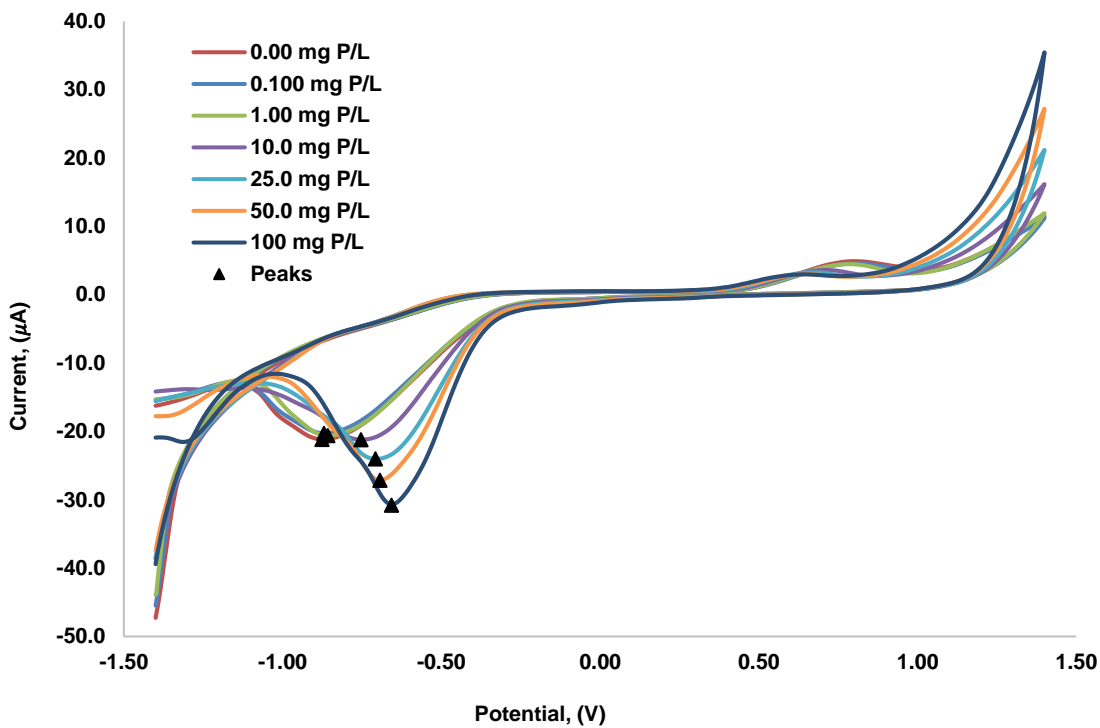

**Figure A-1. Average cyclic voltammograms acquired from experiments conducted on synthetic phosphate solutions at pH 4.00.**

The points in Figure A-1 were regressed against their respective phosphate concentrations to obtain the calibration curves in Figure 3-2 and Figure 3-3 at pH of 4.00. Table A-2 and Figure A-2 summarise the regression's statistical outputs for the calibration curves.

**Table A-2. Regression statistics for synthetic phosphate solution at pH of 4.00 calibration curves acquired using Excel's Data Analysis tool. Table A-2a represents the current versus phosphate concentration curve while Table A-2b represents the potential versus logarithmic phosphate concentration curve.**

| Regression Statistics (a) | | Regression Statistics (b) | |
|---|---|---|---|
| Multiple R | 0.986 | Multiple R | 0.970 |
| R Square | 0.972 | R Square | 0.942 |
| Adjusted R Square | 0.967 | Adjusted R Square | 0.927 |
| Standard Error | 0.729 | Standard Error | 24.0 |
| Observations | 7 | Observations | 6 |

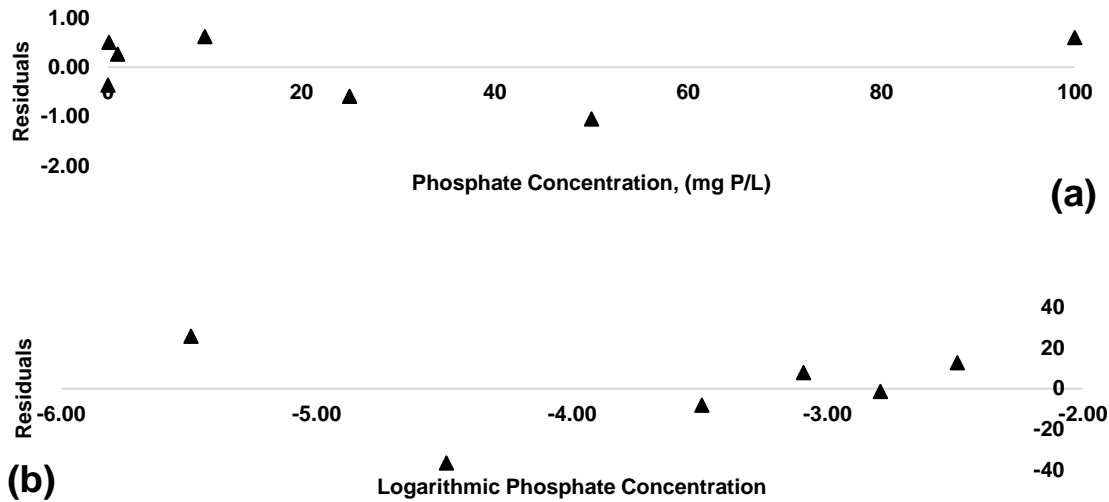

**Figure A-2. The residual plots of the calibration curves of the synthetic phosphate solutions at pH 4.00. Figure (a) represents the current versus phosphate concentration curve while (b) represents the potential versus logarithmic phosphate concentration curve.**

Equation (A-1) represents the correlation between the peak currents and phosphate concentrations.

$$i_{peak} = -(0.12 \pm 0.02) \cdot [P] - (20.8 \pm 0.9)$$ (A-1)

Where,

$i_{peak}$: Peak current in $\mu$A.

$[P]$: Phosphate concentration in mg P/L.

To calculate "the lowest amount of analyte in a sample that can be detected with (stated) probability," as the LOD is defined by the National Committee for Clinical Laboratory Standards (NCCLS), Equation (A-2) was utilised (2004).

$$LOD = \mu_{Blank} + 1.645 \cdot \sigma_{Blank} + 1.645 \cdot \sigma_{Lowest\ [P]} \quad \textbf{(A-2)}$$

Where,

$LOD$: Limit of detection for the current in $\mu$A.

$\mu_{Blank}$: Peak current of the blank in $\mu$A.

$\sigma_{Blank}$: Standard deviation of the peak current of the blank in $\mu$A.

$\sigma_{Lowest\ [P]}$: Standard deviation of the peak current of the lowest phosphate concentrated sample in $\mu$A.

$$LOD = 21.2 + 1.645 \cdot (0.5) + 1.645 \cdot (0.5) = 22.8\ \mu A$$

Equation (A-1) was then used to calculate the corresponding phosphate concentration.

$$-22.8 = -0.12 \cdot [P] - 20.8$$

$$[P] = 18.5\ mg\ P/L^*$$

* The mathematical operation above dose not result in the solution given because the mathematical operation shown shows minimal significant figures, where the actual calculation done on Microsoft Excel was not restrained by such a limit.

The error associated with the LOD was calculated in accordance with guidelines set out by the University of California, Davis (2017). Equation (A-3) was used to calculate the error for calculations involving addition and subtraction and Equation (A-4) was used to calculate the error for calculations involving multiplication and division.

$$\epsilon_x = \sqrt{\epsilon_a^2 + \epsilon_b^2 + \cdots + \epsilon_z^2} \tag{A-3}$$

$$\frac{\epsilon_x}{x} = \sqrt{\left(\frac{\epsilon_a}{a}\right)^2 + \left(\frac{\epsilon_b}{b}\right)^2 + \cdots + \left(\frac{\epsilon_z}{z}\right)^2} \tag{A-4}$$

Where,

$\epsilon$: Error associated with a value. Carries the unit of the parent value.

$x$: The value associated with the error.

$a, b, \ldots, z$: Values involved in calculations.

The error associated with the $LOD$ was carried from the standard deviation of the blank's peak current because no other error affected the calculation of the $LOD$. On the other hand, calculating the corresponding phosphate concentration involved more errors.

$$\epsilon_y = \sqrt{\epsilon_B^2 + \epsilon_{i_{peak}}^2}$$

Where,

$B$: The y-intercept in the calibration Equation (A-1).

$$\epsilon_y = \sqrt{(0.9)^2 + (0.5)^2} = 1.02$$

$$\frac{\epsilon_{[P]}}{[P]} = \sqrt{\left(\frac{\epsilon_y}{y}\right)^2 + \left(\frac{\epsilon_S}{S}\right)^2}$$

Where,

$S$: The slope in the calibration Equation (A-1).

$$\frac{\epsilon_{[P]}}{18.5} = \sqrt{\left(\frac{1.02}{-22.8+20.8}\right)^2 + \left(\frac{0.02}{-0.12}\right)^2}$$

$$\epsilon_{[P]} = 10$$

LOD of the cobalt based sensor at pH of 4.00 in a synthetic phosphate solution:-

$$[P] = (18 \pm 10)\ mg\ P/L$$

## A.2 pH 8.00

Table A-3 and Figure A-3 provide the data points observed in the cyclic voltammograms tested at different phosphate concentrations. For each concentration, 4 data replicates were taken. The average of the 4 readings was then calculated.

**Table A-3. Average raw data acquired from experiments conducted on synthetic phosphate solutions at pH 8.00.**

| Phosphorus Concentration (mg P/L) | Average Peak Current (uA) | Standard Deviation | Average Peak Potential (V) | Standard Deviation |
|---|---|---|---|---|
| 0.00 | -29.7 | 0.8 | -1.09 | 0.02 |
| 0.100 | -29.3 | 0.7 | -1.08 | 0.01 |
| 1.00 | -30 | 1 | -1.04 | 0.01 |
| 10.0 | -33.4 | 0.6 | -0.87 | 0.01 |
| 25.0 | -39 | 2 | -0.828 | 0.006 |
| 50.0 | -46.0 | 0.9 | -0.807 | 0.008 |
| 100 | -57 | 2 | -0.74 | 0.01 |

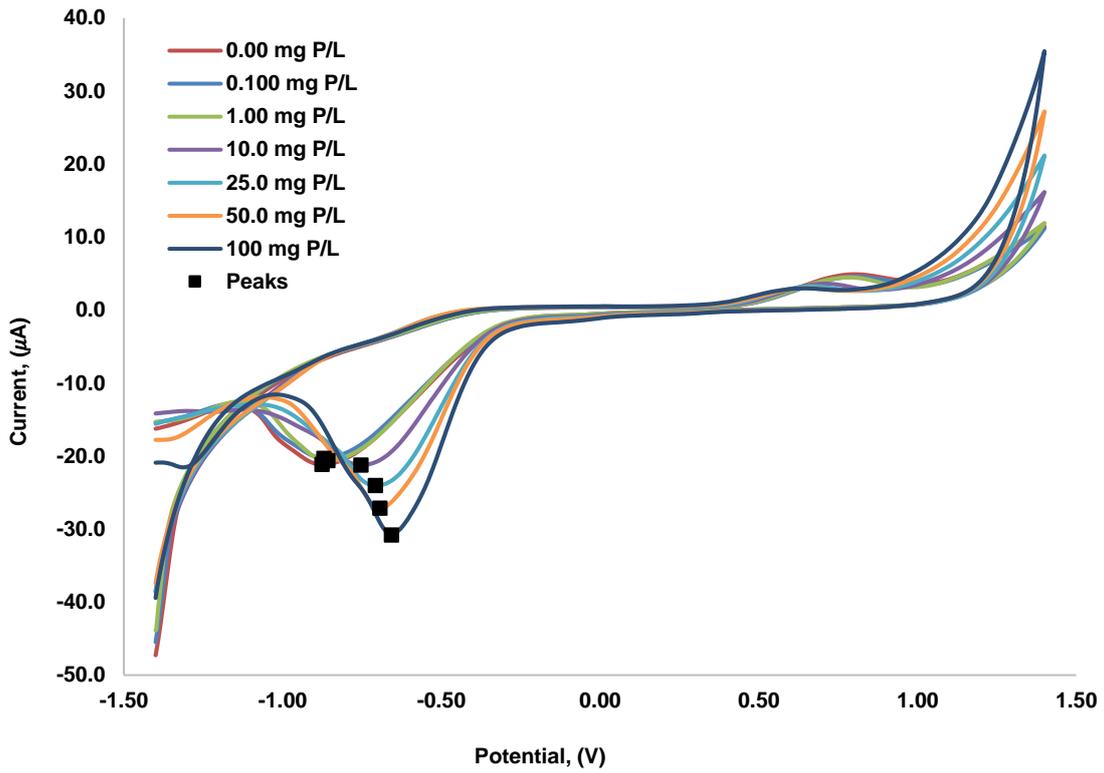

**Figure A-3. Average cyclic voltammograms acquired from experiments conducted on synthetic phosphate solutions at pH 8.00.**

The points in Figure A-3 were regressed against their respective phosphate concentrations to obtain the calibration curves in Figure 3-2 and Figure 3-3 at pH of 8.00. Table A-4 and Figure A-4 summarise the regression's statistical outputs for the calibration curves.

**Table A-4. Regression statistics for synthetic phosphate solution at pH of 8.00 calibration curves acquired using Microsoft Excel's Data Analysis tool. Table A-2a represents the current versus phosphate concentration curve while Table A-2b represents the potential versus logarithmic phosphate concentration curve.**

| Regression Statistics (a) | | Regression Statistics (b) | |
|---|---|---|---|
| Multiple R | 0.990 | Multiple R | 0.980 |
| R Square | 0.981 | R Square | 0.960 |
| Adjusted R Square | 0.977 | Adjusted R Square | 0.950 |
| Standard Error | 1.58 | Standard Error | 30.2 |
| Observations | 7 | Observations | 6 |

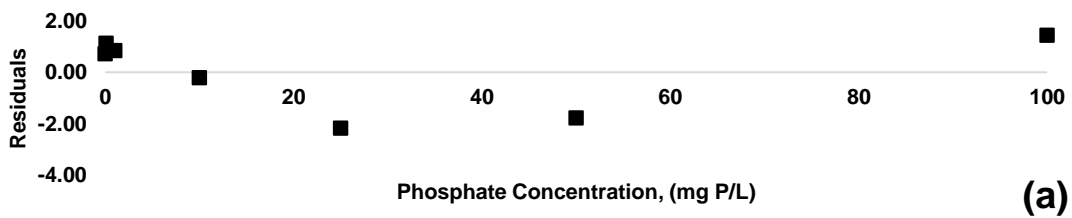

(a)

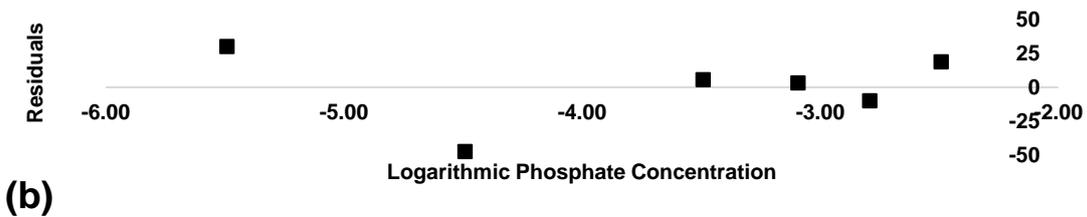

(b)

**Figure A-4. The residual plots of the calibration curves of the synthetic phosphate solutions at pH 8.00. (a) represents the current versus phosphate concentration curve while (b) represents the potential versus logarithmic phosphate concentration curve.**

Equation (A-5) represents the correlation between the peak currents and phosphate concentrations.

$$i_{peak} = -(0.28 \pm 0.04) \cdot [P] - (30 \pm 2)$$ (A-5)

Equations (A-2) - (A-4) were used to calculate the LOD of the cobalt based sensor at pH 8.00 in a synthetic phosphate solution:-

$$[P] = (6 \pm 8)\ mg\ P/L$$

## A.3 Statistical Analysis

A two-way ANOVA statistical test was conducted to corroborate the effect of pH on the sensor's performance. Table A-5 details the raw data used to conduct the statistical analysis. The first null hypothesis (Sample) suggested that there was no significant difference between the current responses provided at different pH. The second null hypothesis (Columns) suggested that there was no significant difference between the current responses provided at different phosphate concentrations. The third null hypothesis (Interaction) suggested that there was a significant difference between the pH and the phosphate concentration in their effect on the current response. The hypotheses were tested by observing the Fisher Factor (F) and P-values provided by the ANOVA tests. If the F<F-Critical and P<0.05, the null hypotheses were accepted. Otherwise, they were rejected.

**Table A-5. Set of peak currents in $\mu$A acquired at pH 4.00 and 8.00 organised under two independent factors; pH and phosphate concertation.**

| Phosphate Concentration (mg P/L) | 0.000 | 0.100 | 1.00 | 10.0 | 25.0 | 50.0 | 100 |
|---|---|---|---|---|---|---|---|
| pH 4.00 | -21.0 | -19.8 | -19.9 | -20.9 | -23.6 | -27.3 | -29.6 |
|  | -21.3 | -20.0 | -19.8 | -20.9 | -24.2 | -26.3 | -31.5 |
|  | -20.6 | -20.8 | -21.3 | -21.7 | -24.0 | -26.9 | -30.1 |
|  | -21.7 | -20.7 | -21.6 | -21.4 | -24.4 | -28.2 | -30.9 |
| pH 8.00 | -29.2 | -30.1 | -29.4 | -34.0 | -36.5 | -45.4 | -54.6 |
|  | -28.8 | -28.6 | -28.0 | -33.5 | -40.1 | -45.5 | -55.4 |
|  | -30.6 | -29.6 | -31.3 | -32.7 | -40.8 | -45.8 | -57.3 |
|  | -30.1 | -28.9 | -30.6 | -33.3 | -40.7 | -47.4 | -59.5 |

Table A-6 represents the output from the two-way ANOVA statistical analysis produced using Microsoft Excel.

**Table A-6. Two-Way ANOVA test conducted on the sensor's response acquired at different pH and phosphate concentrations.**

| Source of Variation | SS | df | MS | F | P-Value | F-Critical |
|---|---|---|---|---|---|---|
| Sample | 2819 | 1 | 2819 | 2421 | 0.00 | 4.07 |
| Columns | 2454 | 6 | 409 | 351 | 0.00 | 2.32 |
| Interaction | 513 | 6 | 85.5 | 73.4 | 0.00 | 2.32 |
| Within | 48.9 | 42 | 1.16 | | | |
| Total | 5835 | 55 | | | | |

Because F>F-Critical and P<0.05 in all sources of variation, all the null hypotheses were rejected and the alternative hypotheses were accepted.

# Appendix B – Effect of Interfering Ions Data Analysis

Appendix B provides the raw data acquired from experiments conducted on synthetic phosphate and interfering ion solutions at different concentrations of phosphate and pH of 8.00. The appendix also provides the cyclic voltammograms produced from the experiments. The appendix also details all the data analysis carried out on the data set.

## B.1 Cl$^-$

Table B-1 and Figure B-1 provide the data points observed in the cyclic voltammograms tested at different phosphate concentrations and 100 mg Cl$^-$/L. For each concentration, 4 data replicates were taken. The average of the 4 readings was then calculated.

**Table B-1 Average raw data acquired from experiments conducted on synthetic phosphate solutions with 100 mg Cl$^-$/L as interfering ion at pH 8.00.**

| Phosphorus Concentration (mg P/L) | Average Peak Current (uA) | Standard Deviation | Average Peak Potential (V) | Standard Deviation |
|---|---|---|---|---|
| 0.00 | -39.1 | 0.6 | -0.733 | 0.003 |
| 0.100 | -39 | 2 | -0.73 | 0.01 |
| 1.00 | -39 | 1 | -0.72 | 0.01 |
| 10.0 | -40 | 2 | -0.71 | 0.01 |
| 25.0 | -42.7 | 0.9 | -0.687 | 0.003 |
| 50.0 | -45 | 1 | -0.682 | 0.004 |
| 100 | -49 | 1 | -0.66 | 0.01 |

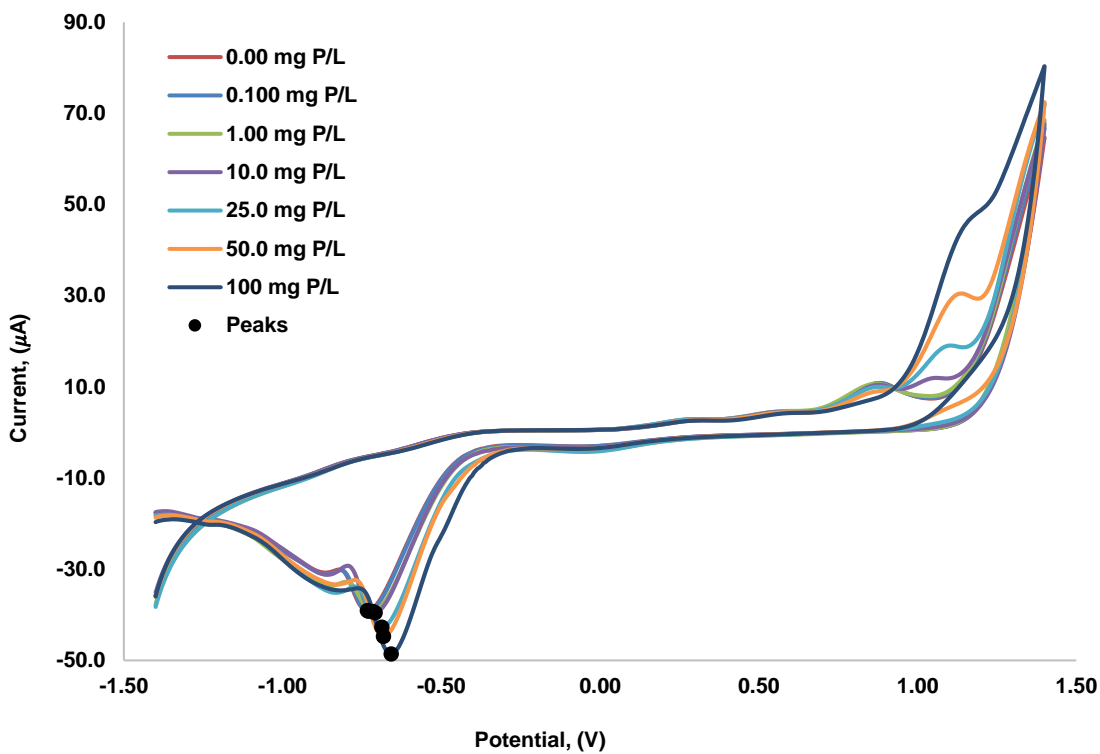

**Figure B-1. Average cyclic voltammograms acquired from experiments conducted on synthetic phosphate solutions with 100 mg Cl⁻/L as interfering ion at pH 8.00.**

The points in Figure B-1 were regressed against their respective phosphate concentrations to obtain the calibration curves in Figure 3-4 at 100 mg Cl⁻/L. Table B-2 and Figure B-2 summarise the regression's statistical outputs for the calibration curves.

**Table B-2. Regression statistics for synthetic phosphate solutions with 100 mg Cl⁻/L as interfering ion at pH 8.00's calibration curve acquired using Microsoft Excel's Data Analysis tool.**

| Regression Statistics | |
|---|---|
| Multiple R | 0.987 |
| R Square | 0.974 |
| Adjusted R Square | 0.969 |
| Standard Error | 0.649 |
| Observations | 7 |

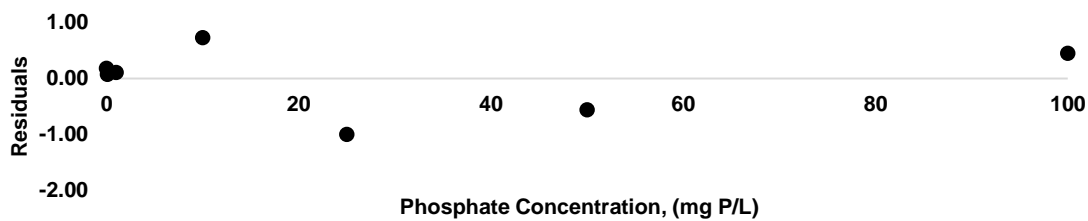

**Figure B-2. The residual plots of the calibration curve of the synthetic phosphate solutions with 100 mg Cl⁻/L as interfering ion at pH 8.00.**

Equation (B-1) represents the correlation between the peak currents and phosphate concentrations.

$$i_{peak} = -(0.10 \pm 0.02) \cdot [P] - (39.3 \pm 0.8) \tag{B-1}$$

## B.2 $SO_4^{2-}$

Table B-3 and Figure B-3 provide the data points observed in the cyclic voltammograms tested at different phosphate concentrations and 100 mg $SO_4^{2-}$/L. For each concentration, 4 data replicates were taken. The average of the 4 readings was then calculated. However, data points at 1.00 mg P/L and 10.0 mg P/L were not considered because they were out of the current response range between 0.00 – 100 mg P/L.

**Table B-3. Average raw data acquired from experiments conducted on synthetic phosphate solutions with 100 mg $SO_4^{2-}$/L as interfering ion at pH 8.00.**

| Phosphorus Concentration (mg P/L) | Average Peak Current (uA) | Standard Deviation | Average Peak Potential (V) | Standard Deviation |
|---|---|---|---|---|
| 0.00 | -45 | 1 | -0.853 | 0.006 |
| 0.100 | -46 | 1 | -0.84 | 0.02 |
| 25.0 | -47.0 | 0.6 | -0.834 | 0.009 |
| 50.0 | -52 | 1 | -0.80 | 0.01 |
| 100 | -57 | 2 | -0.77 | 0.01 |

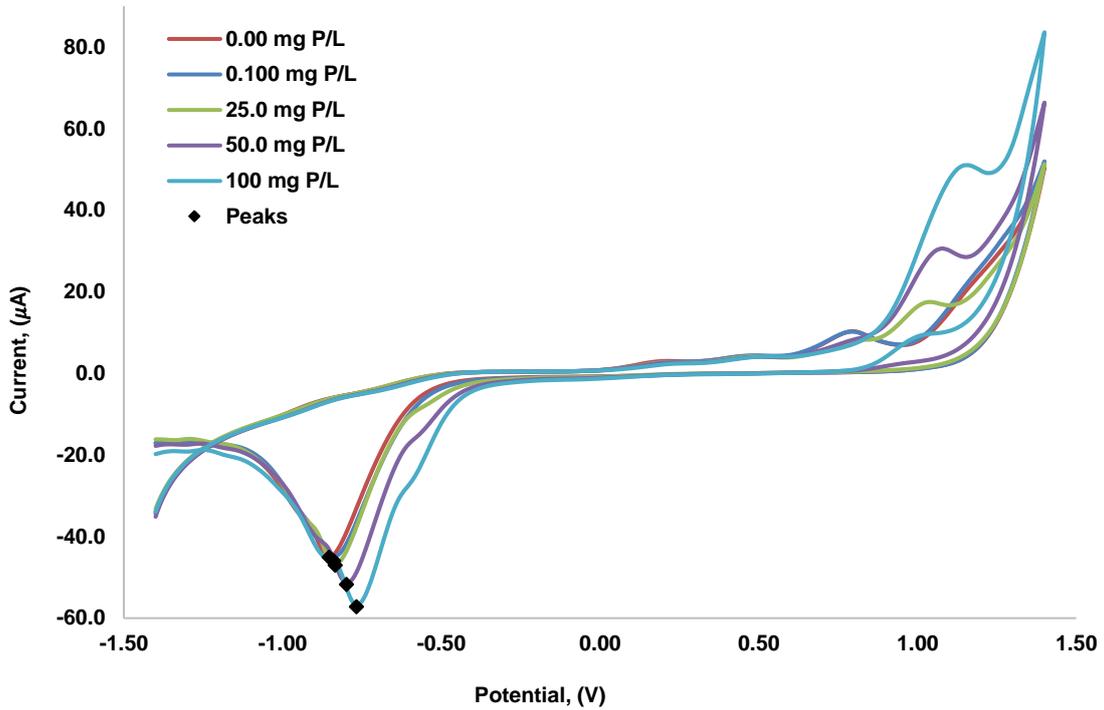

**Figure B-3. Average cyclic voltammograms acquired from experiments conducted on synthetic phosphate solutions with 100 mg $SO_4^{2-}$/L as interfering ion at pH 8.00.**

The points in Figure B-3 were regressed against their respective phosphate concentrations to obtain the calibration curve to predict peak current at 1.00 mg P/L and 10.0 mg P/L. Table B-4 summarises the regression's statistical outputs for the calibration curve.

**Table B-4. Regression statistics for synthetic phosphate solutions with 100 mg $SO_4^{2-}$/L as interfering ion at pH 8.00's calibration curve acquired using Microsoft Excel's Data Analysis tool.**

| Regression Statistics | |
|---|---|
| Multiple R | 0.990 |
| R Square | 0.979 |
| Adjusted R Square | 0.972 |
| Standard Error | 0.844 |
| Observations | 5 |

Equation (B-2) represents the correlation between the peak currents and phosphate concentrations. The equation was used to predict the current response at 1.00 mg P/L and 10.0 mg P/L.

$$i_{peak} = -(0.12 \pm 0.03) \cdot [P] - (45 \pm 2)$$  (B-2)

$$i_{peak} = -0.12 \cdot (1.00) - 45$$

$$i_{peak} = -45.3\ \mu A$$

Equation (A-3) was then used to propagate the error associated with the predicted response.

$$\epsilon_y = \sqrt{\epsilon_B^2 + \epsilon_{i_{peak}}^2}$$

$$\epsilon_y = \sqrt{(0.3)^2 + (2)^2} = 2$$

The predict sensor response at 1.00 mg P/L:-

$$i_{peak} = -(45 \pm 2)\ \mu A$$

The same approach was taken to calculate the response at 10.0 mg P/L:-

$$i_{peak} = -(46 \pm 2)\ \mu A$$

With the predicted responses, the regression was re-run. Table B-5 and Figure B-4 summarise the regression's statistical outputs for the calibration curve.

**Table B-5. Regression statistics for synthetic phosphate solutions with 100 mg SO$_4^{2-}$/L as interfering ion at pH 8.00's calibration curve acquired using Microsoft Excel's Data Analysis tool (with the inclusion of the predicted data at 1.00 and 10.0 mg P/L).**

| Regression Statistics | |
|---|---|
| Multiple R | 0.991 |
| R Square | 0.982 |
| Adjusted R Square | 0.979 |
| Standard Error | 0.653 |
| Observations | 7 |

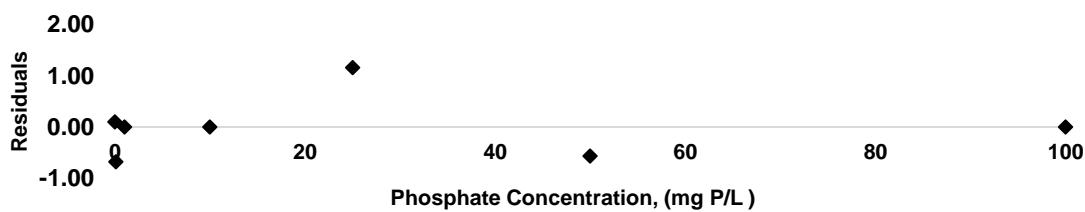

**Figure B-4. The residual plots of the calibration curve of the synthetic phosphate solutions with 100 mg SO$_4^{2-}$/L as interfering ion at pH 8.00.**

Equation (B-3) represents the correlation between the peak currents and phosphate concentrations.

$$i_{peak} = -(0.12 \pm 0.02) \cdot [P] - (45.2 \pm 0.8)$$ (B-3)

## B.3 $NO_3^-$

Table B-6 and Figure B-5 provide the data points observed in the cyclic voltammograms tested at different phosphate concentrations and 100 mg $NO_3^-$/L. For each concentration, 4 data replicates were taken. The average of the 4 readings was then calculated. However, the data point at 10.0 mg P/L was not considered because it was out of current response range between 0.00 and 100 mg P/L.

**Table B-6. Average raw data acquired from experiments conducted on synthetic phosphate solutions with 100 mg $NO_3^-$/L as an interfering ion at pH 8.00.**

| Phosphorus Concentration (mg P/L) | Average Peak Current (uA) | Standard Deviation | Average Peak Potential (V) | Standard Deviation |
|---|---|---|---|---|
| 0.00 | -47 | 1 | -0.879 | 0.008 |
| 0.100 | -47 | 2 | -0.90 | 0.02 |
| 1.00 | -46 | 2 | -0.88 | 0.01 |
| 25.0 | -48 | 1 | -0.835 | 0.009 |
| 50.0 | -52 | 1 | -0.826 | 0.005 |
| 100 | -59 | 2 | -0.798 | 0.003 |

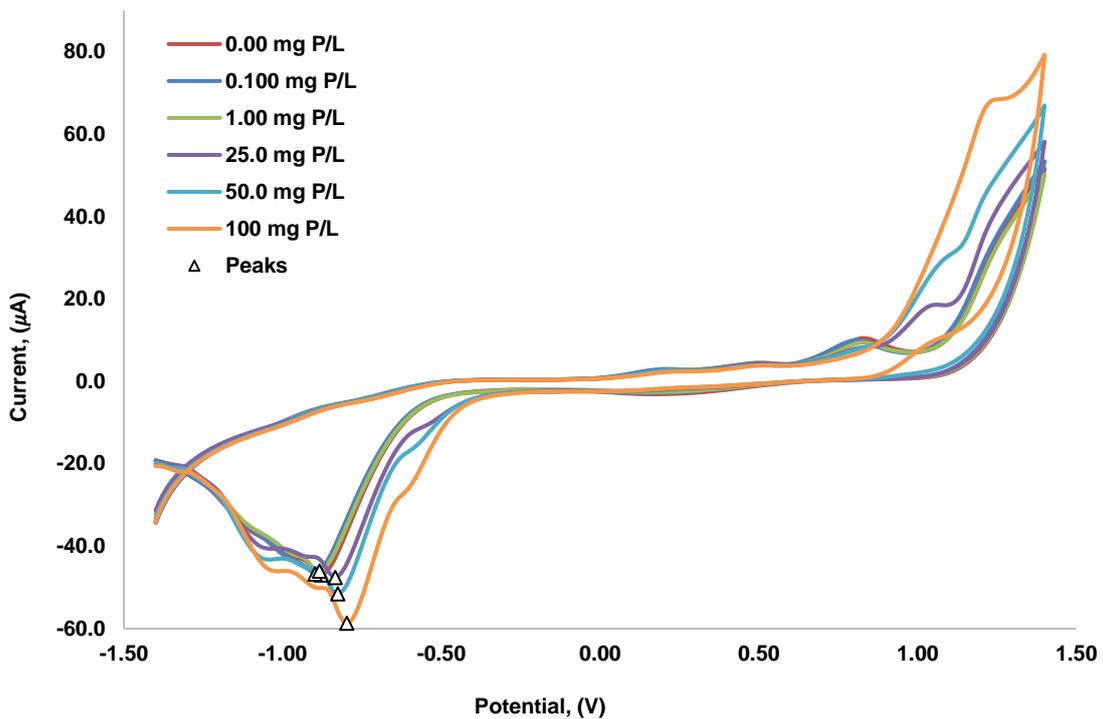

**Figure B-5. Average cyclic voltammograms acquired from experiments conducted on synthetic phosphate solutions with 100 mg $NO_3^-$/L as an interfering ion at pH 8.00.**

The points in Figure B-5 were regressed against their respective phosphate concentrations to obtain the calibration curve to predict the peak current at 10.0 mg P/L. Table B-7 summarises the regression's statistical outputs for the calibration curve.

**Table B-7. Regression statistics for synthetic phosphate solutions with 100 mg $NO_3^-$/L as an interfering ion at pH 8.00's calibration curve acquired using Microsoft Excel's Data Analysis tool.**

| Regression Statistics | |
|---|---|
| Multiple R | 0.982 |
| R Square | 0.965 |
| Adjusted R Square | 0.956 |
| Standard Error | 1.02 |
| Observations | 6 |

Equation (B-4) represents the correlation between the peak currents and phosphate concentrations. The equation was used to predict the current response at 10.0 mg P/L.

$$i_{peak} = -(0.12 \pm 0.03) \cdot [P] - (46 \pm 2)$$ (B-4)

The predicted sensor response at 10.0 mg P/L:-

$$i_{peak} = -(47 \pm 1)\,\mu A$$

With the predicted response, the regression was re-run. Table B-8 and Figure B-6 summarise the regression's statistical outputs for the calibration curve.

**Table B-8. Regression statistics for synthetic phosphate solutions with 100 mg $NO_3^-$/L as an interfering ion at pH 8.00's calibration curve acquired using Microsoft Excel's Data Analysis tool (with the inclusion of the predicted at 10.0 mg P/L).**

| Regression Statistics | |
|---|---|
| Multiple R | 0.983 |
| R Square | 0.966 |
| Adjusted R Square | 0.959 |
| Standard Error | 0.912 |
| Observations | 7 |

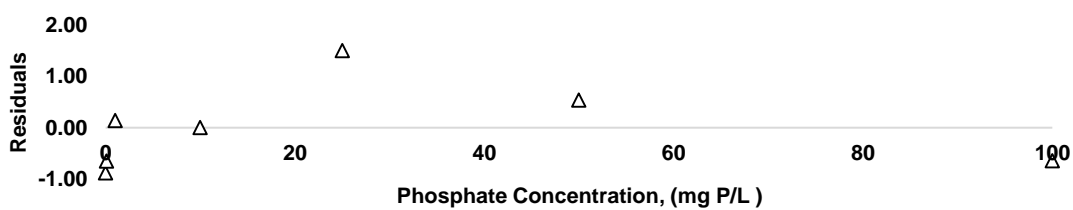

**Figure B-6. The residual plots of the calibration curve of the synthetic phosphate solutions with 100 mg $NO_3^-$/L as an interfering ion at pH 8.00.**

Equation (B-5) represents the correlation between the peak currents and phosphate concentrations.

$$i_{peak} = -(0.12 \pm 0.03) \cdot [P] - (46 \pm 1)$$ **(B-5)**

## B.4 Statistical Analysis

To corroborate the effect of the interfering ions ($Cl^-$, $SO_4^{2-}$ and $NO_3^-$) on the sensor's performance, a two-way ANOVA statistical test was conducted. Tables (B-9) – (B-11) detail the raw data used to conduct the statistical analysis for the three interfering ions. The first null hypothesis (Sample) suggested that there was no significant difference between the current responses provided at different interfering ions' concentrations (0.00 and 100 mg/L for each). The second null hypothesis (Columns) suggested that there was no significant difference between the current responses provided at different phosphate concentrations. The third null hypothesis (Interaction) suggested that there was a significant difference between the different interfering ions' concentrations and the phosphate concentrations in their effect on the current response. The hypotheses were tested as mentioned previously in section A.3.

**Table B-9. Set of peak currents in $\mu$A acquired from phosphate solutions through different concentrations of $Cl^-$ in mg $Cl^-$/L as an interfering ion at different phosphate concentrations.**

| Phosphate Concentration (mg P/L) | | 0.000 | 0.100 | 1.00 | 10.0 | 25.0 | 50.0 | 100 |
|---|---|---|---|---|---|---|---|---|
| $Cl^-$ | 100 mg/L | -39.3 | -40.0 | -38.4 | -37.2 | -43.8 | -44.4 | -47.6 |
| | | -38.9 | -36.9 | -38.8 | -39.2 | -42.9 | -45.8 | -49.9 |
| | | -38.4 | -40.2 | -41.0 | -40.7 | -41.5 | -43.4 | -49.2 |
| | | -39.8 | -39.8 | -39.0 | -41.1 | -42.8 | -45.3 | -47.8 |
| $Cl^-$ | 0.00 mg/L | -29.2 | -30.1 | -29.4 | -34.0 | -36.5 | -45.4 | -54.6 |
| | | -28.8 | -28.6 | -28.0 | -33.5 | -40.1 | -45.5 | -55.4 |
| | | -30.6 | -29.6 | -31.3 | -32.7 | -40.8 | -45.8 | -57.3 |
| | | -30.1 | -28.9 | -30.6 | -33.3 | -40.7 | -47.4 | -59.5 |

**Table B-10. Set of peak currents in $\mu$A acquired from phosphate solutions through different concentrations of $SO_4^{2-}$ in mg $SO_4^{2-}$/L as an interfering ion at different phosphate concentrations.**

| Phosphate Concentration (mg P/L) | | 0.000 | 0.100 | 1.00 | 10.0 | 25.0 | 50.0 | 100 |
|---|---|---|---|---|---|---|---|---|
| $SO_4^{2-}$ | 100 mg/L | -43.7 | -47.2 | -43.7 | -44.7 | -46.5 | -52.5 | -55.8 |
| | | -44.8 | -44.2 | -46.9 | -48.0 | -46.5 | -50.7 | -55.4 |
| | | -44.8 | -46.0 | -43.7 | -44.7 | -47.3 | -52.2 | -58.7 |
| | | -47.1 | -46.1 | -46.9 | -48.0 | -47.8 | -51.6 | -58.8 |
| $SO_4^{2-}$ | 0.00 mg/L | -29.2 | -30.1 | -29.4 | -34.0 | -36.5 | -45.4 | -54.6 |
| | | -28.8 | -28.6 | -28.0 | -33.5 | -40.1 | -45.5 | -55.4 |
| | | -30.6 | -29.6 | -31.3 | -32.7 | -40.8 | -45.8 | -57.3 |
| | | -30.1 | -28.9 | -30.6 | -33.3 | -40.7 | -47.4 | -59.5 |

**Table B-11. Set of peak currents in $\mu$A acquired from phosphate solutions through different concentrations of $NO_3^-$ in mg $NO_3^-$/L as an interfering ion at different phosphate concentrations.**

| Phosphate Concentration (mg P/L) | | 0.000 | 0.100 | 1.00 | 10.0 | 25.0 | 50.0 | 100 |
|---|---|---|---|---|---|---|---|---|
| $NO_3^-$ | 100 mg/L | -48.2 | -48.7 | -43.5 | -45.8 | -48.1 | -52.5 | -55.2 |
| | | -45.7 | -47.7 | -46.4 | -48.8 | -49.2 | -50.1 | -59.1 |
| | | -47.0 | -46.0 | -48.2 | -45.8 | -47.1 | -52.4 | -60.6 |
| | | -47.0 | -44.8 | -46.3 | -48.8 | -46.0 | -51.2 | -59.9 |
| $NO_3^-$ | 0.00 mg/L | -29.2 | -30.1 | -29.4 | -34.0 | -36.5 | -45.4 | -54.6 |
| | | -28.8 | -28.6 | -28.0 | -33.5 | -40.1 | -45.5 | -55.4 |
| | | -30.6 | -29.6 | -31.3 | -32.7 | -40.8 | -45.8 | -57.3 |
| | | -30.1 | -28.9 | -30.6 | -33.3 | -40.7 | -47.4 | -59.5 |

Tables (B-12) – (B-14) represent the output from the two-way ANOVA statistical analysis produced using Microsoft Excel.

**Table B-12. Two-Way ANOVA test conducted on the sensor's response acquired through different Cl⁻ and phosphate concentrations.**

| Source of Variation | SS | df | MS | F | P-Value | F-Critical |
|---|---|---|---|---|---|---|
| Sample | 238 | 1 | 238 | 139 | 0.00 | 4.07 |
| Columns | 2376 | 6 | 396 | 231 | 0.00 | 2.32 |
| Interaction | 547 | 6 | 91.1 | 53.1 | 0.00 | 2.32 |
| Within | 72.0 | 42 | 1.71 | | | |
| Total | 3233 | 55 | | | | |

**Table B-13. Two-Way ANOVA test conducted on the sensor's response acquired through different $SO_4^{2-}$ and phosphate concentrations.**

| Source of Variation | SS | df | MS | F | P-Value | F-Critical |
|---|---|---|---|---|---|---|
| Sample | 1570 | 1 | 1570 | 771 | 0.00 | 4.07 |
| Columns | 2632 | 6 | 439 | 215 | 0.00 | 2.32 |
| Interaction | 449 | 6 | 74.8 | 36.7 | 0.00 | 2.32 |
| Within | 85.6 | 42 | 2.04 | | | |
| Total | 4737 | 55 | | | | |

**Table B-14. Two-Way ANOVA test conducted on the sensor's response acquired through different $NO_3^-$ and phosphate concentrations.**

| Source of Variation | SS | df | MS | F | P-Value | F-Critical |
|---|---|---|---|---|---|---|
| Sample | 1853 | 1 | 1853 | 782 | 0.00 | 4.07 |
| Columns | 2614 | 6 | 436 | 184 | 0.00 | 2.32 |
| Interaction | 472 | 6 | 78.7 | 33.2 | 0.00 | 2.32 |
| Within | 99.5 | 42 | 2.37 | | | |
| Total | 5038 | 55 | | | | |

Because F>F-Critical and P<0.05 in all sources of variation in Tables (B-12) – (B-14), all the null hypotheses were rejected and the alternative hypotheses were accepted.

While the previous set of ANOVA statistical tests evaluated the relation between the interfering ions and the current response in the absence of interfering ions, the next set of ANOVA tests evaluated the relation among the different interfering ions. The first null hypothesis (Sample) suggested that there was no significant difference between the current responses provided at the effect of different interfering ions. The second null hypothesis (Columns) suggested that there was no significant difference between the current responses provided at different phosphate concentrations. The third null hypothesis (Interaction) suggested that there was a significant difference between different interfering ions and different phosphate concentrations in their effect on the current's response. The hypotheses were tested, as mentioned previously in section A.3. Tables (B-15) – (B-18) represent the output from the two-way ANOVA statistical analysis produced using Microsoft Excel.

**Table B-15. Two-Way ANOVA test conducted on the sensor's response acquired through the effect of different interfering ions and different phosphate concentrations.**

| Source of Variation | SS | df | MS | F | P-Value | F-Critical |
|---|---|---|---|---|---|---|
| Sample | 906 | 2 | 453 | 208 | 0.00 | 3.14 |
| Columns | 1262 | 6 | 210 | 96.6 | 0.00 | 2.25 |
| Interaction | 36.6 | 12 | 3.05 | 1.40 | 0.190 | 1.91 |
| Within | 137 | 63 | 2.18 | | | |
| Total | 2342 | 83 | | | | |

**Table B-16. Two-Way ANOVA test conducted on the sensor's response acquired through the effect of $SO_4^{2-}$ and $NO_3^-$ as interfering ions and different phosphate concentrations.**

| Source of Variation | SS | df | MS | F | P-Value | F-Critical |
|---|---|---|---|---|---|---|
| Sample | 11.6 | 1 | 11.6 | 4.66 | 0.04 | 4.07 |
| Columns | 968 | 6 | 161 | 64.8 | 0.00 | 2.32 |
| Interaction | 5.40 | 6 | 0.900 | 0.362 | 0.90 | 2.32 |
| Within | 105 | 42 | 2.49 | | | |
| Total | 1090 | 55 | | | | |

**Table B-17. Two-Way ANOVA test conducted on the sensor's response acquired through the effect of $SO_4^{2-}$ and $Cl^-$ as interfering ions and different phosphate concentrations.**

| Source of Variation | SS | df | MS | F | P-Value | F-Critical |
|---|---|---|---|---|---|---|
| Sample | 586 | 1 | 586 | 319 | 0.00 | 4.07 |
| Columns | 790 | 6 | 132 | 71.7 | 0.00 | 2.32 |
| Interaction | 20.0 | 6 | 3.33 | 1.81 | 0.12 | 2.32 |
| Within | 77.1 | 42 | 1.84 | | | |
| Total | 1473 | 55 | | | | |

**Table B-18. Two-Way ANOVA test conducted on the sensor's response acquired through the effect of $NO_3^-$ and $Cl^-$ as interfering ions and different phosphate concentrations.**

| Source of Variation | SS | df | MS | F | P-Value | F-Critical |
|---|---|---|---|---|---|---|
| Sample | 762 | 1 | 762 | 350 | 0.00 | 4.07 |
| Columns | 785 | 6 | 131 | 60.0 | 0.00 | 2.32 |
| Interaction | 29.4 | 6 | 4.90 | 2.25 | 0.06 | 2.32 |
| Within | 91.6 | 42 | 2.18 | | | |
| Total | 1668 | 55 | | | | |

Because F>F-Critical and P<0.05 in all Sample and Columns sources of variation in Tables (B-15) – (B-18), the first and the second null hypotheses were rejected in all cases and the alternative hypotheses were accepted. On the other hand, the third null hypothesis was accepted for all cases because F<F-Critical and P>0.05 in all Interaction in Tables (B-15) – (B-18).

# Appendix C – Effect of Dissolved Oxygen Data Analysis

      Appendix C provides the raw data acquired from experiments conducted on synthetic phosphate and 1.00 mg $O_2$/L at different concentrations of phosphate and pH of 8.00. The appendix also provides the cyclic voltammograms produced from the experiments. The appendix also details all the data analysis conducted on the data set. Table C-1 and Figure C-1 provide the data points observed in the cyclic voltammograms tested at different phosphate concentrations and 1.00 mg $O_2$/L. For each concentration, 2 data replicates were taken. The average of the 2 readings was then calculated. Only 2 replicates were run because the phosphate solution depleted from oxygen was not stable: the DO levels increased with time because the ambient oxygen in the laboratory dissolved into the synthetic solutions.

**Table C-1. Average raw data acquired from experiments conducted on synthetic phosphate solutions with 1.00 mg $O_2$/L at pH 8.00.**

| Phosphorus Concentration (mg P/L) | Average Peak Current (uA) | Standard Deviation | Average Peak Potential (V) | Standard Deviation |
|---|---|---|---|---|
| 0.00 | -6.2 | 0.2 | -0.860 | 0.01 |
| 0.100 | -6.34 | 2 | -0.86 | 0.01 |
| 1.00 | -6.5 | 0.3 | -0.82 | 0.01 |
| 10.0 | -7.5 | 0.3 | -0.768 | 0.003 |
| 25.0 | -11 | 1 | -0.755 | 0.004 |
| 50.0 | -16 | 1 | -0.718 | 0.006 |
| 100 | -27.8 | 0.4 | -0.72 | 0.02 |

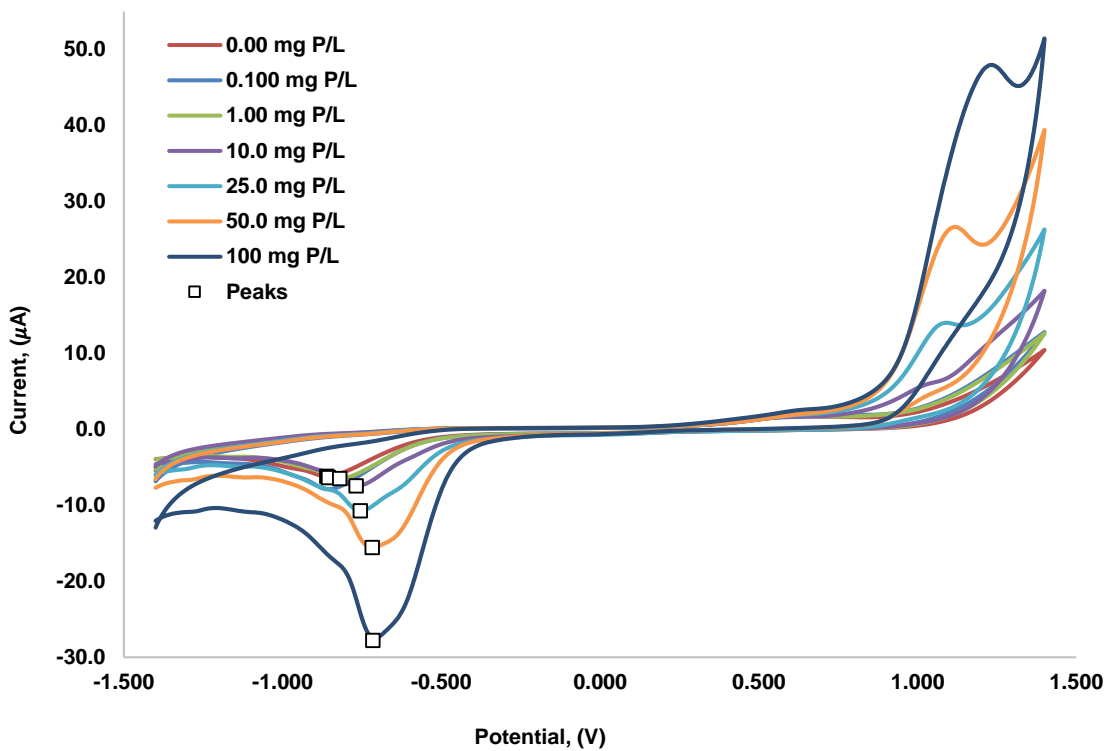

**Figure C-1. Average cyclic voltammograms acquired from experiments conducted on synthetic phosphate solutions with 1.00 mg $O_2$/L at pH 8.00.**

The points in Figure C-1 were regressed against their respective phosphate concentrations to obtain the calibration curves in Figure 3-5 at 1.00 mg $O_2$/L. Table C-2 and Figure C-2 summarise the regression's statistical outputs for the calibration curve.

**Table C-2. Regression statistics for synthetic phosphate solutions with 1.00 mg $O_2$/L at pH 8.00 calibration curve acquired using Microsoft Excel's Data Analysis tool.**

| Regression Statistics | |
|---|---|
| Multiple R | 0.997 |
| R Square | 0.994 |
| Adjusted R Square | 0.993 |
| Standard Error | 0.676 |
| Observations | 7 |

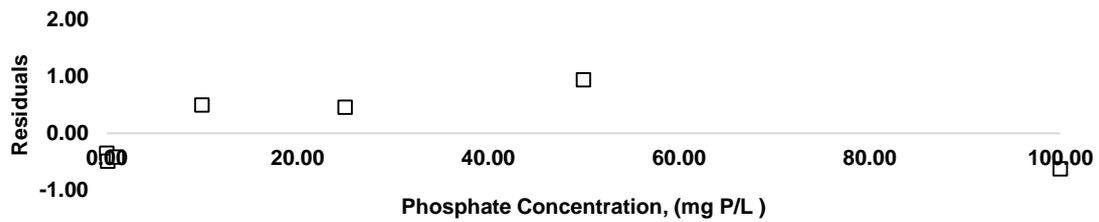

**Figure C-2. The residual plots of the calibration curve of the synthetic phosphate solutions with 1.00 mg $O_2$/L at pH 8.00.**

Equation C-1 represents the correlation between the peak currents and phosphate concentrations.

$$i_{peak} = -(0.21 \pm 0.02) \cdot [P] - (5.8 \pm 0.8) \tag{C-1}$$

## C.1 Statistical Analysis

To corroborate the effect of DO levels on the sensor's performance, a two-way ANOVA statistical test was conducted. Table C-3 details the raw data used to conduct the statistical analysis. The first null hypothesis (Sample) suggested that there was no significant difference between the current responses provided at different DO levels (8.54 and 1.00 mg $O_2$/L). The second null hypothesis (Columns) suggested that there was no significant difference between the current responses provided at different phosphate concentrations. The third null hypothesis (Interaction) suggested that there was a significant difference between the DO levels and the phosphate concentrations in their effect on the current response. The hypotheses were tested as mentioned previously in section A.3.

**Table C-3. Set of peak currents in $\mu$A acquired from phosphate solutions at different concentrations at different DO levels and phosphate concentrations.**

| Phosphate Concentration (mg P/L) | | 0.000 | 0.100 | 1.00 | 10.0 | 25.0 | 50.0 | 100 |
|---|---|---|---|---|---|---|---|---|
| $O_2$ | 1.00 mg/L | -6.05 | -6.34 | -6.28 | -7.70 | -11.3 | -14.8 | -28.1 |
| | | -6.32 | -9.42 | -6.66 | -7.23 | -10.2 | -16.4 | -27.5 |
| | | -6.05 | -6.34 | -6.28 | -7.70 | -11.3 | -14.8 | -28.1 |
| | | -6.32 | -9.42 | -6.66 | -7.23 | -10.2 | -16.4 | -27.5 |
| $O_2$ | 8.54 mg/L | -29.2 | -30.1 | -29.4 | -34.0 | -36.5 | -45.4 | -54.6 |
| | | -28.8 | -28.6 | -28.0 | -33.5 | -40.1 | -45.5 | -55.4 |
| | | -30.6 | -29.6 | -31.3 | -32.7 | -40.8 | -45.8 | -57.3 |
| | | -30.1 | -28.9 | -30.6 | -33.3 | -40.7 | -47.4 | -59.5 |

The experiment conducted on phosphate solutions with 1.00 mg $O_2$/L was replicated twice. However, the experiments involving the synthetic phosphate solution with DO level of 8.54 mg $O_2$/L were replicated 4 times. To conduct the ANOVA test, the same number of replications was required. To fulfil that need, the 2 depleted oxygen readings were duplicated.

Table C-4 represents the output from the two-way ANOVA statistical analysis produced using Microsoft Excel.

**Table C-4. Two-Way ANOVA test conducted on the sensor's response acquired at different DO levels and phosphate concentrations.**

| Source of Variation | SS | df | MS | F | P-Value | F-Critical |
|---|---|---|---|---|---|---|
| Sample | 9500 | 1 | 9500 | 7389 | 0.00 | 4.07 |
| Columns | 3918 | 6 | 653 | 508 | 0.00 | 2.32 |
| Interaction | 141 | 6 | 23.5 | 18.3 | 0.00 | 2.32 |
| Within | 54.0 | 42 | 1.29 | | | |
| | | | | | | |
| Total | 13613 | 55 | | | | |

Because F>F-Critical and P<0.05 in all sources of variation, all the null hypotheses were rejected and the alternative hypotheses were accepted.

# Appendix D - Tests on Real Wastewater Samples Data Analysis

Appendix D provides the raw data acquired from experiments conducted on real wastewater samples and the associated data analysis.

Equation A-5 was used to translate the peak currents acquired from real wastewater samples. However, the current peak for the effluent samples was out of the range of responses acquired from synthetic phosphate model solutions used to produce Equation A-5 with. Therefore, the equation will not be applicable.

Activated sludge mixed liquors sample:-

$$i_{peak} = -0.28 \cdot [P] - 30$$

$$-33 = -0.28 \cdot [P] - 30$$

$$[P] = 10.8 \; mgP/L$$

To estimate the error associated with the predicted phosphate concentration above, Equations A-3 and A-4 were used.

$$\epsilon_y = \sqrt{\epsilon_B^2 + \epsilon_{i_{peak}}^2}$$

$$\epsilon_y = \sqrt{(1.94)^2 + (1.68)^2} = 2.57$$

$$\frac{\epsilon_{[P]}}{[P]} = \sqrt{\left(\frac{\epsilon_y}{y}\right)^2 + \left(\frac{\epsilon_S}{S}\right)^2}$$

$$\frac{\epsilon_{[P]}}{10.8} = \sqrt{\left(\frac{2.57}{-33+30}\right)^2 + \left(\frac{0.04}{-0.28}\right)^2}$$

$$\epsilon_{[P]} = 9$$

The phosphate concentration of the mixed liquors samples acquired from sensors using the calibration curve at pH 8.00:-

$$[P]_{ML} = (11 \pm 9) \; mg \; P/L$$

For effluent samples, the same procedure shown above was utilised. The phosphate concentration of the effluent samples acquired from sensors using the calibration curve at pH 8.00:-

$$[P]_{FE} = (58 \pm 12)\ mg\ P/L$$

## D.1 Wastewater Characterisation

**Table D-1. Extended wastewater characterisation results.**

| Component Name | Results | | | Units |
|---|---|---|---|---|
| | Influent | Mixed Liquors | Effluent | |
| Temperature | 23.8 | 22.7 | 22.5 | °C |
| pH | 7.60 | 7.00 | 7.10 | - |
| DO | 0.900 | 5.99 | 8.50 | mg $O_2$/L |
| Phosphate | 3.87 | 0.210 | 0.160 | mg P/L |
| $Cl^-$ | 97.5 | 89.2 | 91.0 | mg $Cl^-$/L |
| $SO_4^{2-}$ | 73.4 | 125 | 124 | mg $SO_4^{2-}$/L |
| $NO_3^-$ | 0.885 | 87.2 | 105 | mg $NO_3^-$/L |
| Biochemical Oxygen Demand | 60.7 | 5.30 | <1.90 | mg $O_2$/L |
| Chemical Oxygen Demand | 122 | 57.5 | 26.3 | mg $O_2$/L |
| Alkalinity | 379 | 129 | 108 | mg $CaCO_3$/L |
| Ammoniacal Nitrogen | 33.6 | 0.820 | <0.0200 | mg N/L |
| Nitrite | <0.00800 | 0.100 | 0.032 | mg N/L |
| Nitrogen Total Oxidised | <0.200 | 19.8 | 23.7 | mg N/L |
| Aluminium | 56.2 | 26.6 | 10.2 | µg/L |
| Barium | 15.0 | 4.30 | 5.10 | µg/L |
| Beryllium | <0.200 | <0.200 | <0.200 | µg/L |
| Boron | 91.0 | 85.0 | 88.0 | µg/L |

| Component Name | Results | | | Units |
|---|---|---|---|---|
| | Influent | Mixed Liquors | Effluent | |
| Calcium | 104 | 94.1 | 93.6 | mg/L |
| Chromium | <0.900 | <0.900 | <0.900 | µg/L |
| Copper | 25.0 | 4.00 | <3.00 | µg/L |
| Iron | 1380 | 1060 | 78.4 | µg/L |
| Lithium | 11.6 | 11.0 | 10.8 | µg/L |
| Magnesium | 5.90 | 6.60 | 6.20 | mg/L |
| Manganese | 42.9 | 184 | 48.3 | µg/L |
| Molybdenum | 3.60 | 3.40 | 2.80 | µg/L |
| Potassium | 15.9 | 19.4 | 17.8 | mg/L |
| Sodium | 64.2 | 71.4 | 71.2 | mg/L |
| Strontium | 0.318 | 0.271 | 0.277 | mg/L |
| Zinc | 43.0 | 9.00 | 8.00 | µg/L |

## D.2 Statistical Analysis

To corroborate that the phosphate concentration obtained from real wastewater samples using the cobalt based sensor was statistically similar to the readings from the wastewater characterisation, a two-tailed t-test was conducted; Equation D-1.

$$t = \frac{\bar{X} - \mu_o}{S/\sqrt{n}} \qquad \text{(D-1)}$$

Where,

$t$: Calculated t-value.

$\bar{X}$: The phosphate concentration acquired from real wastewater sample in mg P/L.

$\mu_o$: The phosphate concentration acquired from wastewater characterisation in mg P/L.

$\mu_o$: The standard deviation of the phosphate concentration acquired from real wastewater sample in mg P/L.

$n$: The number of replicates taken to acquire reading from real wastewater sample.

The null hypothesis stated that the two values were the same, while the alternative hypothesis stated that they were not. The two hypotheses were evaluated by calculating a t-value that had to be lower than the prescribed t-values obtained from Student's t-tables for the null hypothesis to be accepted.

Mixed liquor samples:-

$$t = \frac{10.8 - 0.210}{8.57/\sqrt{3}}$$

$$t = 2.15$$

Effluent samples:-

$$t = \frac{57.6 - 0.160}{12.1/\sqrt{3}}$$

$$t = 8.23$$

Student's t-test:-

$$t_{0.05,3} = 3.1824$$

For mixed liquor samples, because $t < t_{0.05,3}$, the null hypothesis was accepted. On the other hand, for effluent samples, because $t > t_{0.05,3}$, the null hypothesis was rejected and the alternative hypothesis was accepted.**)**